\documentclass[prl,twocolumn,aps,superscriptaddress]{revtex4-2}

\usepackage{graphicx}

\usepackage{amsmath,amssymb}

\usepackage{dcolumn}
\usepackage{bm}
\usepackage{amsthm}
\usepackage[shortlabels]{enumitem}

\usepackage{mathtools} 

\usepackage{graphicx}
\usepackage{makecell} 
\usepackage{longtable}
\usepackage{array}
\usepackage{booktabs}
\usepackage{tabularx}
\usepackage{accents}   

\theoremstyle{plain}
\newtheorem{theorem}{Theorem}[section]

\newtheorem{proposition}[theorem]{Proposition}

\usepackage{thmtools}
\usepackage[dvipsnames]{xcolor} 

\declaretheoremstyle[
    bodyfont=\normalfont,
    headpunct={.}, 
    postheadspace=1em,
]{mydefstyle}

\theoremstyle{remark}

\newcommand{\mycomment}[1]{} 

\usepackage{hyperref}
\hypersetup{hypertexnames=false}

\begin{document}

\preprint{APS/123-QED}

\title{Optimal Proteome Allocation in a Self-Replicating System By Internal Sensing}

\author{Jonathan Shao-Kai Huang}
\affiliation{ 
Department of Physics, National Taiwan University, Taipei 106, Taiwan
}%

\author{Wei-Hsiang Lin}
\affiliation{ 
Institute of Molecular Biology, Academia Sinica, Taipei 115, Taiwan
}%
\affiliation{ 
Department of Physics, National Taiwan University, Taipei 106, Taiwan
}

\date{\today}

\begin{abstract}
Bacteria must divide their finite resources between proteins that take up nutrients and ribosomes that make new proteins. How can a cell find the best balance as nutrient availability varies, and can this be done by sensing partial information of its internal state? In a bacterial growth model, we derive the allocation that maximizes steady-state growth in each environment. We show that, when knowledge of its surroundings is absent, a single rule reaches each optimum by sensing only the internal amino acid pool, and then generalize to a class of multi-signal strategies. This is connected to how the signaling molecule ppGpp could help implement this rule. The model also predicts a crossover in the steady-state growth law, where the optimal ribosome fraction rises as the square root of growth rate under severe nutrient limitation and nearly linearly at intermediate-to-fast growth.
  
\end{abstract}

\maketitle

\emph{Introduction.}---
Biological systems must allocate finite biosynthetic resources among competing processes~\cite{Scott2010,Chure_eLife}, and maximize survival under different environmental and physiological conditions~\cite{khonsari_perception_2015,Giordano2016,Dourado2020AnalyticalBalancedGrowth,towbin_optimality_2017}. In a self-replicating system, often modeled by a chemical reaction network, one of these sectors are dedicated to growth, and is crucial to the study of the allocation strategy.

\begin{figure}[htbp]
    \centering
    \includegraphics[width=0.48\textwidth]{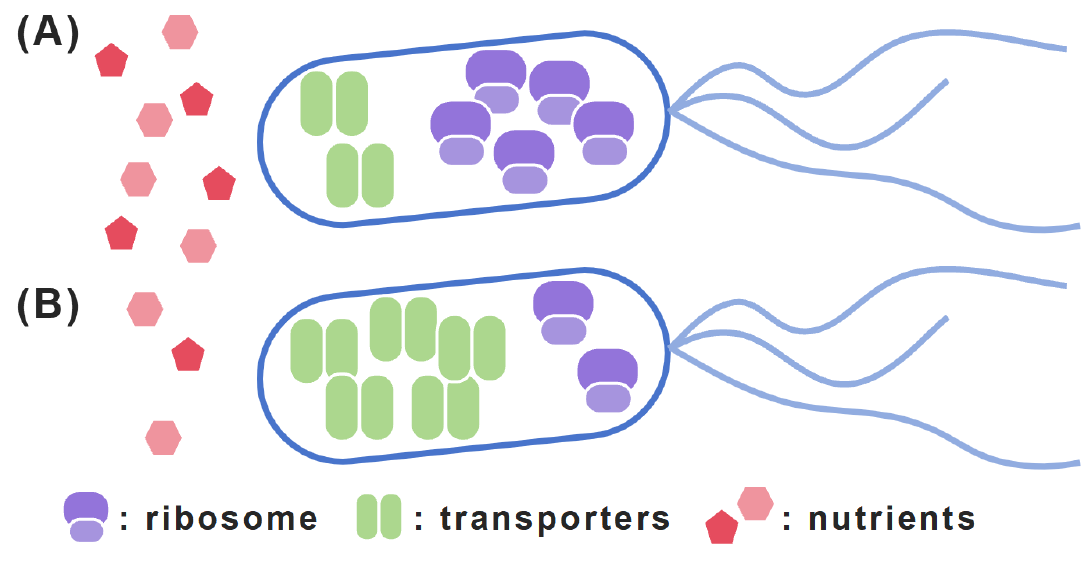}
    \caption{Bacteria have a finite resource pool. (A) In rich nutrient, e.g. glucose, more resource is allocated to the production of ribosomes and the self-replicating machinery. (B) In poor nutrient, e.g. ribose or glycerol, more resource is allocated to the production of transporters and the metabolic machinery.}
    \label{fig:0}
\end{figure}

Microbial growth provides a particularly illuminating setting\cite{molenaar_shifts_2009,scott_bacterial_2011}.
Proteomics and quantitative modeling have revealed groups of proteins that respond coherently to growth modulation~\cite{Hui2015GlobalResourceAllocation,mori_coarse_2021,schmidt_quantitative_2016,Lalanne2021TranslationFactors,mateus_functional_2020,hu_protein_2020,belliveau_fundamental_2021}.
In Figure~\ref{fig:0}, consider a bacterium placed in different nutrient environments. In rich nutrient, it does not need to spend much effort for metabolism, and can allocate more resource to the production of ribosomes, the replication machinery; in poor nutrient, however, it needs to express more transporters and metabolic machinery just to achieve flux balance.
Extensive experiments beginning in the late 1950s~\cite{schaechter_dependency_1958,neidhardt_studies_1960,kjeldgaard_distribution_1963,davison_molecular_1966,maaloe_control_1966,ecker_synthesis_1969} till recently~\cite{nissen_flux_1997,binder_growth_1998,worden_growth_2003,bremer_modulation_2008,nicklisch_rnaprotein_2009,Scott2014,dinh_evaluating_2023,gao_yeast_2025} have also revealed strikingly robust steady-state relationships between steady-state growth rate and proteome allocation, known as \emph{growth laws}~\cite{Scott2010}. 

For a system with externally prescribed \emph{fixed allocation}, the analysis is straightforward as a scalar optimization problem. Yet, more commonly, biological regulation operates only through a small number of internal signals~\cite{You2013,wu_cellular_2022,scott_shaping_2023}. This produces a nonlinear feedback system with \emph{adaptive allocation}, which can be studied as a robust optimal-control problem~\cite{Giordano2016,yegorov_optimal_2019,yabo_dynamical_2022,Chure_eLife,innerarity_imizcoz_optimal_2024,imizcoz_optimal_2025}. 

To explain these observations, fine-grained computational models have been proposed~\cite{karr_whole_cell_2012,Weiss2015Mechanistic,goelzer_quantitative_2015,mori_constrained_2016}, while coarse-grained models with a small number of sectors~\cite{Scott2010,Scott2014,bosdriesz_how_2015,maitra_bacterial_2015,Hui2015GlobalResourceAllocation,Baldazzi2023,Droghetti2026,lai_coarse_grained_2026} and low-dimensional models have been studied~\cite{Giordano2016,Pandey2016AnalyticGrowthLaws,Dourado2020AnalyticalBalancedGrowth,Chure_eLife,seeger_enzyme_2026}. Notably, models inspired by (p)ppGpp regulation~\cite{liao_how_2024} that approximate optimal allocation have found popularity~\cite{Sharma2018ProteomeRegulation,Chure_eLife}. These works establish resource allocation as a powerful organizing principle of microbial physiology. Moreover, understanding cellular allocation has practical consequences for metabolic engineering~\cite{teusink_resource_2026} and antibiotic research~\cite{greulich_growthdependent_2015,salzer_role_2023}.  

We derive the optimal strategy for a proteome partition model at steady-state. Growth laws emerge as an intermediate regime of a nonlinear curve, consistent with empirical studies. We then characterize adaptive strategies that dynamically regulate toward optimality, state various criteria for convergence, and propose connections to the ppGpp regulation mechanism.

\emph{Model.}---
Consider a generic biological system described by a state vector $\mathcal{X}(t) \in \mathbb{R}_{\geq 0}^{n}$. The dynamics of $\mathcal{X}$ depend on an allocation vector $\phi \in \Delta_{m-1}$, where $\Delta_{m-1}$ is the $(m-1)$-simplex. The components of $\phi$ represent the fractions of a collective resource allocated to competing processes. The dynamics are assumed to take the \emph{autonomous} form $\frac{\mathrm{d}\mathcal{X}}{\mathrm{d}t} = F\left(\mathcal{X},\phi\,|\,\eta\right)$, where $\eta$ denotes \emph{environmental parameters} such as nutrient quality. When clear from context, $\eta$ may be left out.

Biological regulatory systems generally do not have complete access to $\eta$. Instead, they often sense internally accessible signals, such as metabolite concentration, tRNA charge, or signaling molecules\cite{You2013,Kochanowski2013,wu_cellular_2022,Chure_eLife}. We assume that regulation with \emph{adaptive allocation} acts through a reduced set of physiological observables.
We write the adaptive \emph{strategy} as $\phi = \phi(\mathcal{X}(t))$, and which may depend only on some components of $\mathcal{X}(t)$.
When the system converges stably to a fixed point $\mathcal{X}^*$, freezing $\phi \equiv \phi(\mathcal{X}^*)$ defines the corresponding \emph{fixed allocation}.

\begin{figure*}[htbp]
    \centering
    \includegraphics[width=0.95\textwidth]{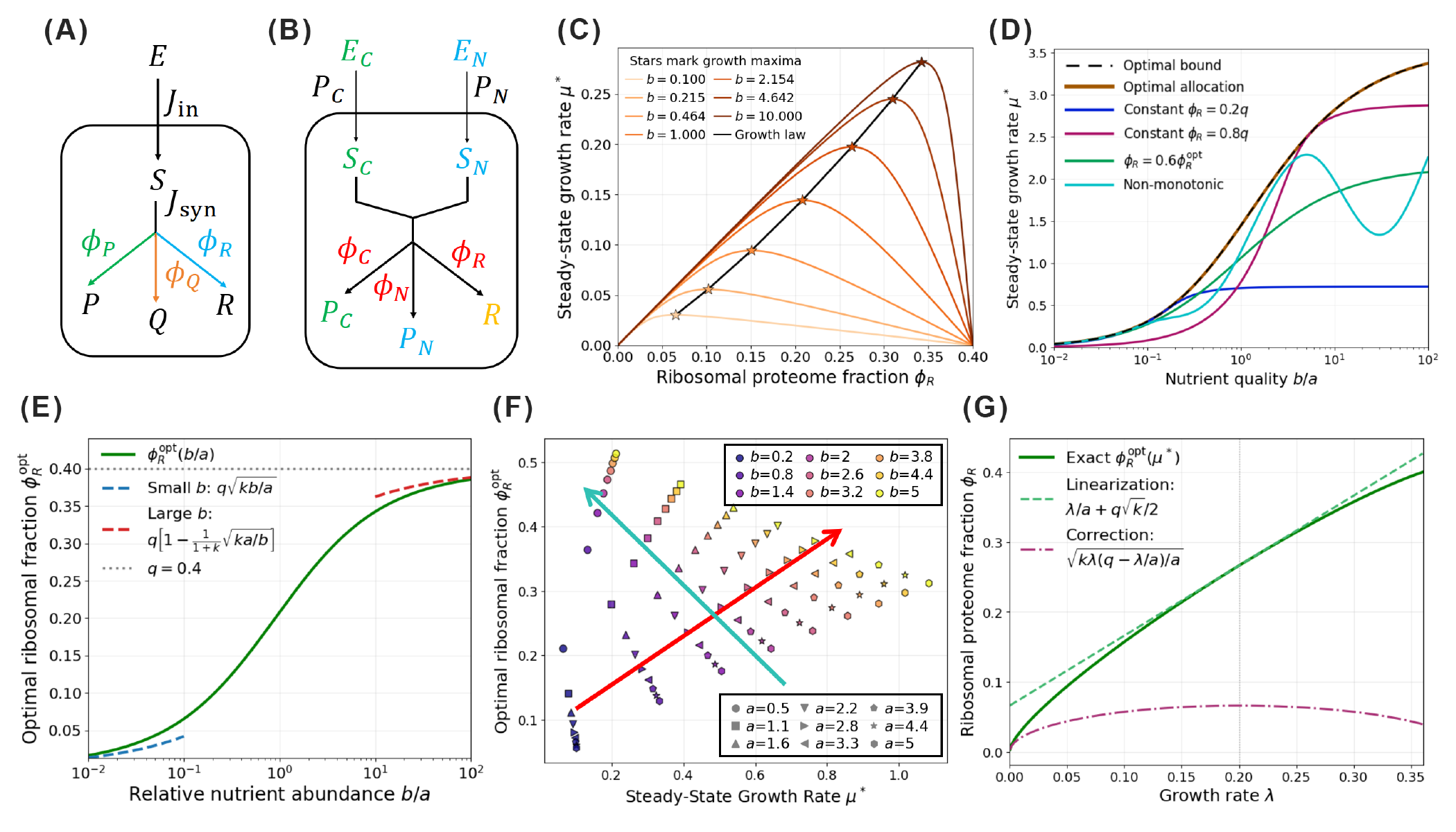}
    \caption{(A) When the strategy depends on internal signals only, i.e. $\phi_R = \phi_R(S,P,Q,R)$, this is the adaptive three-sector model. When the strategy depends on the environment, i.e. $\phi_R=\phi_R(a,b)$, this is the constant-allocation three-sector model. (B) Complementary carbon-nitrogen resource model, motivated by coarse-grained models of parallel nutrient uptake and joint biomass synthesis~\cite{kleijn_coarse_grained_2023}. (C) Steady-state growth rate $\mu^*$ attains a unique maximum at $\phi_R^{\rm opt}$. (D) Optimal allocation $\phi_R^{\rm opt}$ as a function of nutrient quality $b$. When $b/a\ll 1$, low ribosome allocation $\phi_R=0.2q$ approximates the optimum; when $b/a\gg 1$, high ribosome allocation $\phi_R=0.8q$ approximates the optimum.Growth rate under any strategy is bounded by the optimal growth rate $\mu^{\rm opt}$.
    (E) Optimal fixed-allocation strategy for the constant proteome partition model. The poor- and rich-medium asymptotes are shown for $b/a < 10^{-1}$ and $b/a>10^1$, respectively. (F) Simulation of 81 nutrient and translational perturbations. Red shows growth law under nutrient quality modulation. Cyan shows growth law under translational inhibition. (G) (Green) Exact first growth law curve with $q=0.4$ and $k=0.1$. (Light green) Linearization about $\lambda = qa/2$.}
    \label{fig:1}
\end{figure*}

A bacterium, or more generally a growing system, can be described by a self-replicating system~\cite{molenaar_shifts_2009,Giordano2016,kostinski_ribosome_2020,roy_unifying_2021}. We consider the proteome partition model in Figure~\ref{fig:1}(A), which has the adaptive strategy $\phi_R (S,P,Q,R)$. This nonlinear model is difficult to solve in general and leads to robust behavior (SI). To gain analytic insight, we study the corresponding constant-allocation model, where the strategy is constrained to be independent of dynamics. This is called the \emph{proteome partition model}.

Let $[S], [P], [Q], [R]$ describe the metabolite and amino acid pool, transporter, housekeeping, and ribosomal protein concentrations, respectively. Also, let $b$ be the nutrient abundance, $a$ the maximal protein-synthesis rate, and $k$ the half-saturation constant for ribosome-catalyzed protein synthesis. Estimate parameter values are $a\sim 8$--$12\,\mathrm{h}^{-1}$, $b\sim 0.1$--$10\,\mathrm{h}^{-1}$, and $k\sim 0.003$--$0.03$. In \textit{E. coli}, available proteome fraction is $\phi_Q\sim0.5$. We write
\begin{subequations}
    \begin{align}
        \frac{\mathrm{d} [S]}{\mathrm{d} t} &= b[P] - \frac{a[S][R]}{k+[S]}-\mu[S] ,\\
        \frac{\mathrm{d} [P]}{\mathrm{d} t} &= \phi_P \left(\frac{a [S][R]}{k + [S]}\right) - \mu [P] ,\\
        \frac{\mathrm{d} [Q]}{\mathrm{d} t} &= \phi_Q \left(\frac{a [S][R]}{k+[S]}\right) - \mu [Q], \\
        \frac{\mathrm{d} [R]}{\mathrm{d} t} &= \phi_R \left(\frac{a [S][R]}{k + [S]}\right) - \mu [R],
    \end{align}
\end{subequations}
where the specific growth rate is $\mu = b[P]/\rho$ by flux-balance and mass conservation. Let the biomass be $X= (X_S, X_P, X_Q, X_R)$, and $M$ their sum. Then we define \emph{composition} as the biomass fraction $Y = X/M$, such that the cell density $\rho$ satisfies $[X] = \rho Y$. 

This is a \emph{scalable reaction network (SRN)}~\cite{Lin2020}, where the vector field $F$ such that $\dot{X} = F(X)$ is homogeneous of degree one, i.e. $ F(\alpha X) = \alpha F(X)$ for all $\alpha > 0$ and $X \in \mathbb{R}_{\ge 0}^m $. While not every exponentially growing system is an SRN, since $\rho$ is approximately constant across balanced-growth conditions~\cite{MartinezSalas1981,Basan2015,Pang2023}, SRN is a natural setting for cell models. Furthermore, since a \emph{concentration} equation and a \emph{composition} equation differ only by a conversion factor $\rho$, from here on we adopt the latter notation.

\begin{figure*}[htbp]
    \centering
    \includegraphics[width=0.95\textwidth]{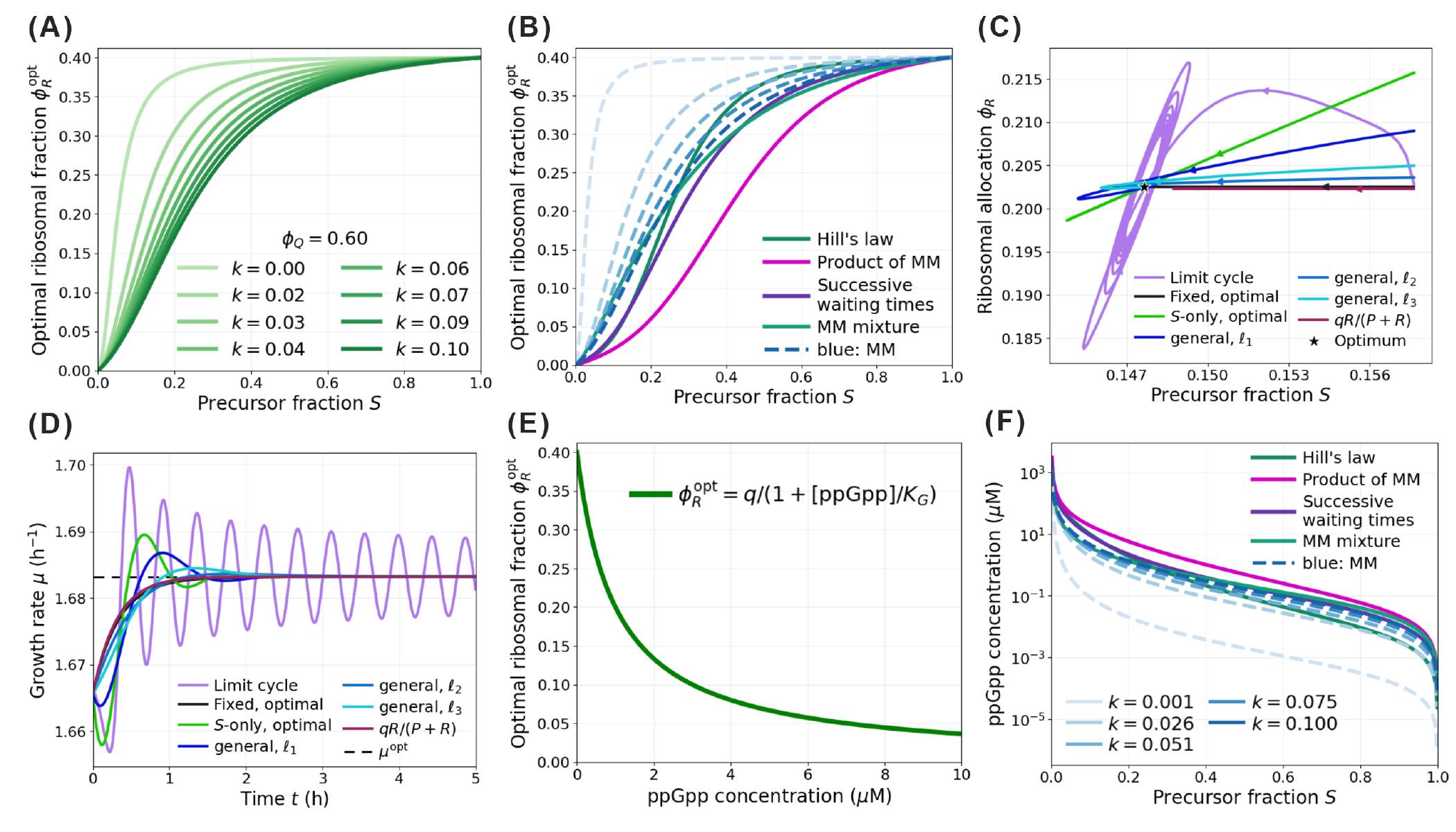}
    \caption{(A) Optimal adaptive allocation strategy as a function of precursor fraction $S$, for 10 values of $k$. (B) Optimal adaptive allocation strategy for models with different $\kappa(S)$: Hill's law, product and sum of Michaelis-Menten reactions, single Michaelis-Menten reaction, and a sum of successive reaction waiting times. (C) Interdependence of $\phi_R$ and amino acid pool fraction $S$ evolving through time. Seven adaptive allocation strategies are considered: fixed $\phi_R^{\rm opt}(b/a)$, single-signal optimal strategy $\psi_R^{\rm opt}(S)$, difference-measurer $\phi_R^{\mathrm{opt}, \ell}(Y)$ with $\ell_1(x)=0.5x$, $\ell_2(x)=0.1x$, $\ell_3(x)=0.2x/(0.5+x)$, strong-feedback with limit cycle, and $qR/(P+R)$. Details of the strategies are in the SI. (D) Same strategies are considered. Damped oscillations are observed for all strategies except the fixed optimal strategy, the suboptimal $qR/(P+R)$, and strong-feedback strategy. (E) Empirical relationship between $\phi_R^{\rm opt}$ and ppGpp concentration, $K_G=1 \,\mathrm{\mu M}$~\cite{bosdriesz_how_2015,Giordano2016}. (F) Theoretically predicted amino acid dependence of ppGpp concentration, according to different $\kappa(S)$.}
    \label{fig:2}
\end{figure*}

Growth rate is a dynamical quantity that characterizes the speed of cell proliferation. Assuming the system reaches a steady state (a nontrivial result to be shown later), the steady-state growth rate $\mu^*$ under fixed proteome partition $\phi_R$ is
\begin{equation}
    \mu^*(\phi_R) = a \phi_R \left(\dfrac{\sqrt{c^2 + 4k}-c}{\sqrt{c^2+4k}-c+2k}\right),
    \label{eq:two_sector_lambda_full}
\end{equation}
where $c \equiv \frac{a\phi_R}{b(q-\phi_R)}+k-1$ and $q\equiv 1-\phi_Q$. 
The steady-state growth rate $\mu^*$ is a good measurement of fitness~\cite{dekel_optimality_2005,towbin_optimality_2017,Dourado2020AnalyticalBalancedGrowth}.
As Figure~\ref{fig:1}(D) shows, for each environment $\eta=(a,b)$, $\mu^*$ attains a environment-specific unique optimum $\mu^{\rm opt}$ at some $\phi_R\in(0,q)$. The fixed-allocation optimal strategy is $\phi_R^{\rm opt} = \operatorname{arg\,max}_{\phi_R} \mu^*(\phi_R)$, and is given by (SI)
\begin{equation}
    \phi_R^{\text{opt}} 
    = 
    q \left[\frac{
        \sqrt{\frac{kb}{a}}+(1+k)\left(\frac{b}{a}\right)
        }{
        1+2\sqrt{\frac{kb}{a}}+(1+k)\left(\frac{b}{a}\right)
        }
    \right]. 
    \label{eq:theta_opt_full}
\end{equation}
The strategy is shown in Figure~\ref{fig:2}(A), with asymptotes in the slow- and fast-growth regimes. This demonstrates an exactly solvable model. Next, we will demonstrate steady-state growth laws, propose strategies for dynamic adaptation, and study their convergence.

\emph{\label{ssec:growth_laws}Emergence of Growth Laws.}---
The canonical model is consistent with observations for bacterial growth laws under \emph{steady state}, a population-level phenomenon~\cite{taniguchi_quantifying_2010,pavlou_single_cell_2025}. In this section, we demonstrate that our coarse-grained model reproduces the growth laws and give an explanation for the emergence of linearity.
Flux nonlinearity plays an important role, for example, growth law under nutrient modulation with mass-action kinetics is $\phi_R^{\rm opt}\propto\sqrt{\lambda}$, where we use the long-term growth rate $\lambda$ in place of $\mu^*$ for simplicity (SI).

Figure~\ref{fig:1}(G) shows the predicted first law. Two qualitative, physically admissible regions can be observed: (I) a concave curve when $b/a\ll 1$ and (II) a linear region with vanishing nonlinear correction. A third reclining region is precluded since $\phi_R^{\rm opt}>q$. 

\emph{Nutrient modulation}~\cite{Scott2010} results in a linear relationship with positive slope, exactly given by 
\begin{equation}
    \phi_{R}^{\rm opt} = \frac{\lambda}{a} + \sqrt{\frac{k\lambda}{a} \left(q - \frac{\lambda}{a}\right)},
\end{equation}
where $ 0 < \lambda < qa/(1+k) $, as in Figure~\ref{fig:2}(C). The nonlinear correction term is a squareroot of a quadratic centered at $ \lambda = qb/2 $, whose correction to slope vanishes near this region. For the plotted $k=0.1$, over $qa/4\le\lambda\le3qa/4$, the curve differs from its central tangent by at most $0.021q$, indicating linearity. Similarly, \emph{translation inhibition}~\cite{Scott2010} results in a linear relationship with negative slope, with 
\begin{equation}
    \phi_{R}^{\rm opt} = q - \frac{1}{1+k} \left[\frac{\lambda}{b} + \sqrt{\frac{k\lambda}{b} \left( q(1+k) - \frac{\lambda}{b}\right)} \right], 
    \label{eq:second_law_housekeeping}
\end{equation}
where $ 0 < \lambda < qb $, as in Figure~\ref{fig:2}(D). The correction to slope vanishes near $ \lambda = qb(1+k)/2 $ with small deviation from the linear approximation. A quasi-linear intermediate region also emerges between $\kappa(S)R$ and $\lambda$. In the $\phi_R^{\rm opt}\to 0$ limit, experimental works show a concave curve~\cite{wang_environment_specificity_2022} as opposed to our concave curve. This may be an effect of protein degradation and ribosome activity~\cite{calabrese_protein_2022,droghetti_incoherent_2025}



\emph{\label{ssec:strategies}Adaptive Allocation.}---
Cells may evolve a dynamical strategy that does not rely on knowledge of $\eta$, but instead senses internal signals such as amino acid concentrations~\cite{hauryliuk_recent_2015,wu_cellular_2022,Sharma2018ProteomeRegulation}. 
We assume the allocation strategy depends on $(S,P,Q,R)$ and not on $(a,b)$, and consider
\begin{subequations}
    \begin{align}
        \frac{\mathrm{d} S}{\mathrm{d} t} &= b P - \frac{a S R}{k + S} - b P S ,\\
        \frac{\mathrm{d} P}{\mathrm{d} t} &= \phi_P(S,P,Q,R) \left(\frac{a S R}{k + S}\right) - b P^2 ,\\
        \frac{\mathrm{d} Q}{\mathrm{d} t} &= \phi_Q \left(\frac{a SR}{k+S}\right) - bPQ, \\
        \frac{\mathrm{d} R}{\mathrm{d} t} &= \phi_R(S,P,Q,R) \left(\frac{a SR}{k + S}\right) - b P R. 
    \end{align}
\end{subequations}

Remarkably, convergence is guaranteed for any well-behaved $S$-only strategy. When restricted to a single precursor signal, the adaptive strategy
\begin{equation}
    \label{eq:psi_opt}
    \psi_R^{\rm opt} (S) = q \left[\frac{S(k+S)}{k + S^2}\right]
\end{equation}
is unique. 
For an adaptive strategy $\phi_R (Y(t)) \in (0,q)$, convergence to the optimal allocation is guaranteed by (1) trajectories $Y(t)$ stay in the interior and (2) metabolite feedback is positive ($\partial_S \phi_R\ge 0$) and ribosomal feedback is bounded ($\partial_v \phi_R< 1$), where $v=qR/(P+R)$. We state the theorems and their proofs in the End Matter. 

Given a model with protein synthesis flux $\kappa(S)$, while the optimal fixed strategy is unique, the optimal adaptive strategy need not be. We propose a class of biology-inspired strategies satisfying the above assumptions: For a monotone response function $\ell$ such that 
$0<\ell'(\cdot)\le 1$,
\begin{equation}
    \label{eq:general_epsilon}
    \phi_{R}^{\rm opt, [\ell]} (Y) = \frac{qR}{P+R} + \ell\left(\psi_R^{\rm opt}(S)\right) - \ell\left(\frac{qR}{P+R}\right)
\end{equation}
converges optimally while exhibiting distinct transients, as in Figure~\ref{fig:2} (C). We also observe relaxation oscillations in the dynamics of $\phi_R^{\rm opt, [\ell]}$ and $\psi_R^{\rm opt}$ as shown in Figure~\ref{fig:2} (D), consistent with experimental measurements under nutrient upshift and downshift~\cite{nguyen_distinct_2021,droghetti_incoherent_2025}.
$qR/(P+R)$ is the currently implemented allocation, while the $\ell$ terms measure deviation from the optimum.

\emph{\label{ssec:ppgpp}Mechanistic Origin of Adaptive Strategies.}---
Mechanistic origin of proteome fine-tuning in \textit{E. coli} has been pinpointed to ppGpp-mediated regulation. The strategy $\psi^{\rm opt}_R(S)$ can be interpreted if we assume an empirical law \(\phi_R=q/(1+[\text{ppGpp}]/K_G)\), as in Figure~\ref{fig:2}(E). For more fine-grained description of protein synthesis $\kappa(S)$ under log-concavity assumptions, for example as (1) a sum of $N$ Michaelis-Menten reactions $\sum_{i=1}^{N}\frac{a_i S}{k_i+S}$ or (2) an activated translation model, we find a similar trend of [ppGpp] dependence on metabolite concentration. The activated translation model is described in Appendix C. Figure~\ref{fig:2}(F) demonstrates this for various log-concave biosynthetic kinetics. Therefore, the precise kinetics we employ in our coarse-grained model is not essential for the qualitative behavior of the optimal strategy.

Then, optimal allocation with Eq.~\eqref{eq:psi_opt} requires that $[\text{ppGpp}]/K_G=k(1-S)/[S(k+S)]$. This combines a deficiency-to-sufficiency ratio with an additional factor \((1+S/k)^{-1}\), suggesting a connection to  tRNA charging. In the $S\ll k$ limit, it reduces to $[\text{ppGpp}]\propto S^{-1}$.

\emph{Information Limitation.}---
Biological reaction network are usually not exactly solvable. Figure~\ref{fig:1}(B) shows a \emph{complementary-resource} (CR) cell model growing on carbon and nitrogen~\cite{kleijn_coarse_grained_2023}, where one reduced set of internal signals is not sufficient, while another of similar dimension is. 
Let $ J \propto \left(\frac{S_C}{k_C + S_C}\right)\left(\frac{S_N}{k_N + S_N}\right)R.  $ describe protein synthesis. The influxes are $b_\beta P_\beta$ and $\eta = (b_\beta, a)$, for $\beta=C,N$. The composition $(S_C, S_N, P_C, P_N, R)$ with allocation $(\phi_C, \phi_N, \phi_R)$ and $(\nu_C, \nu_N)\in\Delta_1$ satisfies 
\begin{subequations}
    \begin{align}
        \dot{S}_{\beta} &= b_{\beta} P_{\beta} - \nu_\beta J - \mu S_{\beta}, \\
        \dot{P}_{\beta} &= \phi_{\beta} J - \mu P_{\beta}, \\
        \dot{R} &= \phi_R J_{\text{syn}} - \mu R,
    \end{align}
\end{subequations}
where $\mu = b_C P_C + b_N P_N$. We can find two environments $\eta, \bar{\eta}$ such that $S_C^{\text{opt}} = \bar{S}_C^{\text{opt}}$ and $P_C^{\text{opt}} = \bar{P}_C^{\text{opt}}$, but $S_N^{\text{opt}} \neq \bar{S}_N^{\text{opt}}$. The corresponding optimal allocations are $\phi_R^{\rm opt}\approx0.263$ and $0.350$, respectively. Hence, no strategy $\phi_R^{\text{opt}}(S_C, P_C)$ could distinguish $\eta$ from $\bar{\eta}$. However, the set $(S_C,S_N)$ suffices (SI).

Therefore, for a reduced set of signals, an optimal strategy $\phi_R^{\text{opt}}$ \emph{need not exist} in principle. 
In the proteome partition model, $S$ is \emph{control-sufficient}~\cite{gu_observations_2025}; in the CR model, $(S_C, P_C)$ is not control-sufficient, while $(S_C,S_N)$ is. The framework applies to allocation problems beyond cellular growth or even biology, whenever actions must be decided based on partially observed states (see SI).




\emph{Conclusion.}---
Biological regulation is realized dynamically through sensing the internal state, a process that can be described by an adaptive strategy. To study the adaptive system, we first study the corresponding fixed system. For the proteome partition model, we derived aenvironment-dependent fixed strategy. This leads to the emergence of growth laws as a crossover regime at moderate-to-fast growth. 

Our dynamical formulation naturally extends to out-of-steady-state behaviors. When restricted to sensing the precursor pool, we derived the unique optimal adaptive strategy; for multi-signal strategies, we characterized a class of smooth difference-measuring strategies and demonstrated their convergence under time evolution. Based on empirical observations of the ppGpp-amino acid (charged tRNA) interdependence, we propose an interpretation for this strategy. Interestingly, these results are not contingent on the specific model of protein synthesis; indeed, convergence and optimality hold for a wide class of biosynthetic kinetics under mild constraints.

A promising direction is shifting environments and suboptimal allocation, which has been extensively studied experimentally~\cite{koch_vivo_1971,brunschede_establishment_1977,mori_quantifying_2017,erickson_global_2017,korem_kohanim_bacterial_2018,balakrishnan_suboptimal_2021,zhu_stringent_2023}. 
This provides a way to test our sensing framework away from equilibrium. More generally, incorporating sensing noise, imperfect responses, and non-deterministic environmental fluctuations allow a more detailed study of dynamic allocation. 

\bibliography{apssamp}

\newpage
\appendix

\begin{figure}[htbp]
    \centering
    \includegraphics[width=0.48\textwidth]{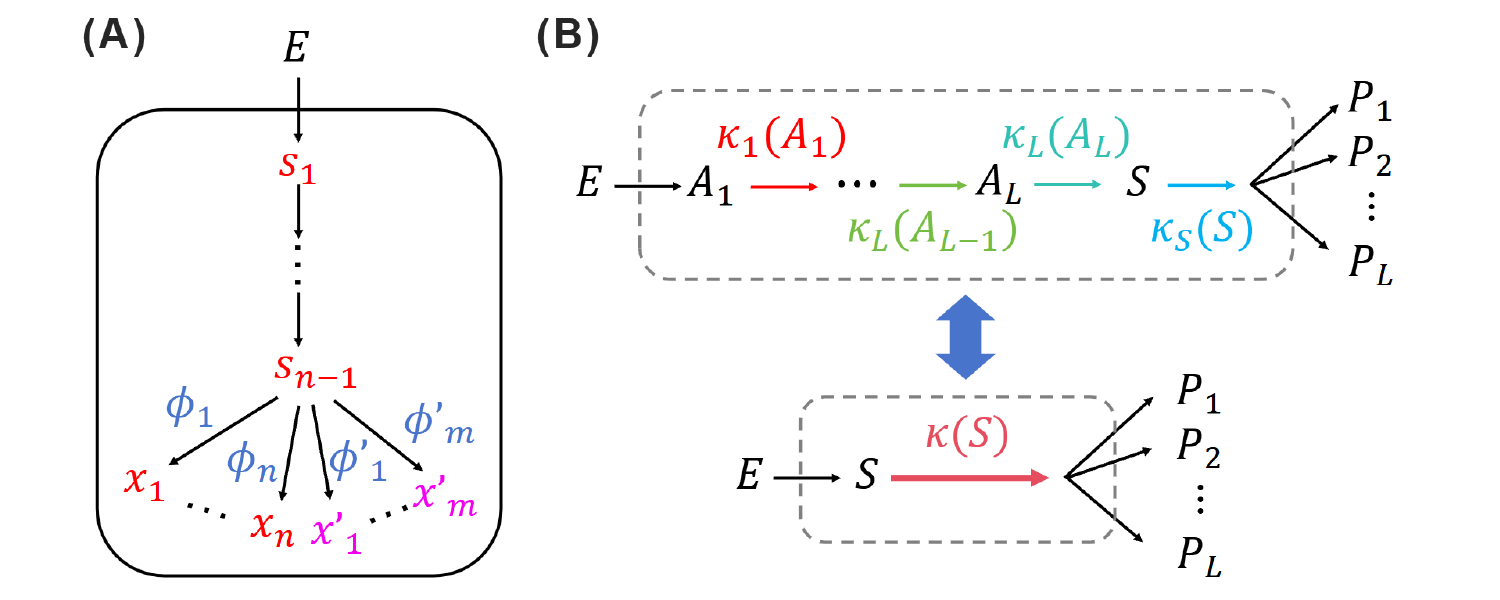}
    \caption{(A) Extended fixed-allocation model with $m+n$ sectors. (B) Coarse-graining the complex protein syntesis process results in complicated expressions of $\kappa(S)$.}
    \label{fig:end}
\end{figure}

\emph{Appendix A: Strategy in Poor-Nutrient Environments.}---
Unlike the canonical model, most nonlinear reaction network models are not exactly solvable. 
The canonical model gives $\phi_R \simeq q\sqrt{kb/a}$ in poor nutrient $b/a\ll 1$. We find an analogue law for the \emph{extended $ (n+m) $-sector models} in Figure~\ref{fig:end}-(A). Here, Michaelis-Menten-type fluxes are replaced with fluxes $R_j$ satisfying the \emph{scalable flux assumption}: (1) $ R_j = R_j (Y_j) $ is scalable; (2) $ R_j^{\prime} (Y_j) > 0 $ for all $ Y_j \geq 0 $. The dynamics are
\begin{subequations}
    \begin{align}
        &\frac{\mathrm{d}Y_1}{\mathrm{d}t} = b Y_n - R_1 (Y_1) Y_{n+1} - b Y_1 Y_n, \\
        &\quad \vdots \notag \\
        &\frac{\mathrm{d}Y_{n-1}}{\mathrm{d}t} = R_{n-2} (Y_{n-2}) Y_{2n-2} \notag \\\
        &\quad\quad\quad - R_{n-1} (Y_{n-1}) Y_{2n-1} - b Y_{n-1} Y_n, \\
        &\frac{\mathrm{d}Y_n}{\mathrm{d}t} = \phi_1 R_{n-1} (Y_{n-1}) Y_{2n-1} - b Y_n^2, \\
        &\frac{\mathrm{d}Y_{n+1}}{\mathrm{d}t} = \phi_2 R_{n-1} (Y_{n-1}) Y_{2n-1} - b Y_n Y_{n+1}, \\
        &\quad \vdots \notag \\
        &\frac{\mathrm{d}Y_{2n-1}}{\mathrm{d}t} = \phi_n R_{n-1} (Y_{n-1}) Y_{2n-1} - b Y_n Y_{2n-1}, \\
        &\frac{\mathrm{d}Y_{2n}}{\mathrm{d}t} = \phi^{\prime}_1 R_{n-1} (Y_{n-1}) Y_{2n-1} - b Y_n Y_{2n}, \\
        &\quad \vdots \notag \\ 
        &\frac{\mathrm{d}Y_{2n+m-1}}{\mathrm{d}t} = \phi^{\prime}_m R_{n-1} (Y_{n-1}) Y_{2n-1} \notag \\
        &\quad\quad\quad - b Y_n Y_{2n+m-1}.
    \end{align}
\end{subequations}
The quantity $(\phi_1, \dots, \phi_n, \phi^{\prime}_1, \dots, \phi^{\prime}_m) \in \Delta_{n+m-1}^{\circ} $ is the \emph{proteome partition fraction} for the constant-allocation system, where $\phi_i$ are functional sectors while $\phi^{\prime}_j$ are housekeeping or gratuitous sectors. A general class of scalable fluxes given by $ J_j (Y) = R_j (Y_j) Y_{j+n} $, for $ j=1,\dots, n-1 $, is considered, where the specific rate functions $ R_j $ are functions of $ Y_j $ only.

Expand $R_j (Y_j) = r_j Y_j + \cdots$, where $ r_j = R_j^{\prime} (0) $. The poor-nutrient optimal strategy is
\begin{equation}
    \phi_{2\le j\le n}^{\rm opt} \simeq \left(1 - \sum_{k=1}^{m} \phi^{\prime}_k\right)\sqrt{\frac{b}{r_{j-1}}},
\end{equation}
with $\phi_1^{\rm opt} \simeq 1 - \sum_{k=1}^{m} \phi^{\prime}_k - \sum_{j=2}^n \phi_j^{\rm opt}$. In particular, the square-root allocation rules are universal modulo the scalable flux assumption. In the $b \ll a$ limit, the relationship $\phi^{\rm opt}_n (b)$ is in general never linear. Hence, the steady-state growth law curve $\lambda(\phi^{\rm opt}_n)$ is also not linear and furthermoe convex.

\newpage
\emph{Appendix B: Proof Sketch.}---

\emph{Definition 1: Persistence}---
A trajectory $\mathcal{X}(t) \subseteq \Omega$ is called \emph{persistent} if it stays in the interior $\Omega^{\circ}$ of $\Omega$ for all $t\in[0,\infty]$, whenever it starts out in $\Omega^{\circ}$. Biologically, this means the system never allocates ``nothing'' to a particular compartment.

\emph{Theorem 1: Convergence}---
Let $v=qR/(P+R)$, $w=(P+R)/(1-S)$, and let $\phi_R(S,v,w)$ be a strategy that is not necessarily optimal. If the trajectories of the adaptive system $Y(t)$ are persistent, and $F(S,v) = \phi_R|_{w=q}$ satisfies $\partial_S F \ge 0$ and $\partial_v F < 1$, then the system converges to the unique balanced-growth state. This state maximizes growth when the feedback law reproduces the optimal allocation at the optimal composition.
\begin{proof}
    Proof sketch: 
    we show $w\to q$ by persistence, so the $\omega$-limit set $\Omega$ of a trajectory lies in the 2D manifold $\{w=q\}$.
    The Poincar\'{e}-Bendixson Theorem~\cite{thieme_convergence_1992} dictates that $\Omega$ must be one of three possibilities: (1) fixed point, (2) periodic orbits, or (3) fixed points connected by heteroclinic or homoclinic orbits. (2) and (3) are precluded per the Dulac criterion whenever $\partial_S F\ge 0$ and $\partial_v F<1$. These conditions also guarantee $H'>0$, $\operatorname{tr} J < 0$, and $\det J>0$ at the fixed point, where $H \equiv F(S,r_0(S)) - r_0(S)$ and $J$ is the Jacobian. Therefore, the $\Omega$ consists of a unique stable fixed point, so the system much reach it. 
\end{proof}

The above theorem provides a sufficient criterion for convergence of admissible adaptive strategies. 

\emph{Theorem 2}---
Suppose $Y(0)\in\Delta_3^{\circ}$, $\ell\in C^1([0,q])$ and $0 < \ell'\le 1$. Then, the class of strategies
\begin{equation}
    \phi^{\rm opt, \ell}_R = v + \ell(\psi(S)) - \ell(v)
\end{equation}
converges asymptotically stably to the the unique optimal allocation $Y^{\rm opt}$.

\begin{proof}
    We show (1) admissibility: $0<\phi_R<q$; (2) persistence: $(S,v,w)$, and hence $(S,P,Q,R)$, stay bounded away from the boundary; and (3) convergence: apply Theorem 1. 
    
    Admissibility follows from the Mean Value Theorem. For persistence, consider the Lyapunov function $W = \frac{q}{w}E_S(S) + E_v(v)$, where
    \begin{subequations}
        \begin{align}
            E_S &= \int_{S^*}^S\mathrm{d}s\, \frac{\kappa(s)}{G(s)}\left[\varepsilon(\psi(s))-\varepsilon(v^*)\right], \\
            E_v &= \int_{v^*}^v\mathrm{d}x\,\left(1-\frac{v^*}{x}\right),
        \end{align}
    \end{subequations}
    and $G(S) = (1-S)\left[b(1-S)+\kappa(S)\right]$. 

    Write down the dynamical equations for $S,v,w$, we show $w\to q$ monotonically and $\int_0^\infty\mathrm{d}t\,|\dot w|\le |w(0)-q|$, and conclude $0 < w_- \le w \le w_+ < q$. 
    Then calculate $\dot W = - \kappa(S)D(S,v) - \frac{\dot w}{w}W_S$.
    Then, $\dot W \le C|\dot w|W$, and hence a well-known inequality implies $W(t)$ is bounded.

    This shows that $(S,v,w)$ is contained in a compact subset of $(0,1)\times(0,q)\times(0,1)$, and thus $(S,P,Q,R)$ is contained in a compact subset of $\Delta_3^{\circ}$, hence persistence. 

    Finally, let $F(S,v) = \phi_R|_{w=q}$, we check that $\partial_S F > 0$ and $\partial_v F < 1$, so Theorem 1 applies.
\end{proof}

\emph{Appendix C: Coarse-Graining the Protein Synthesis Flux.}---
Protein synthesis consists of a complex network of reactions, including transcription, translation, and post-translational modifications. This is represented by a single Michaelis-Menten flux $aSR/(k+S)$, allowing us to capture the essential dynamics. We propose the equivalent coarse-grained result with a general log-concave flux $\kappa(S)R$ sastifying $\kappa>0$, $\kappa'>0$, and $(\log\kappa)''<0$. 

For example, activated translation can be described by $\kappa(S) = \frac{aS}{k+S}\left[f_0 + (1-f_0)\frac{S}{K_{\rm a}+S}\right]$, and provides a biologically motivated example of log-concave biosynthetic kinetics, where $0<f_0<1$ is the activation fraction. Example simulations are provided in the SI.
This represents substrate-dependent elongation with by substrate-dependent ribosome activity.

The optimal adaptive strategy
\begin{equation}
    \psi_R^{\mathrm{opt},\kappa}(S) = q\left[1 + (1-S)(\log\kappa(S))'\right]^{-1}
\end{equation}
converges for any log-concave flux $\kappa(S)$ (SI). 
Matching $\phi_R^{\mathrm{opt},\kappa}$ to the empirical law $\phi_R([\text{ppGpp}])$ gives $[\text{ppGpp}]/K_G = (1-S)(\log\kappa(S))'(S)$, which decreases with amino acid concentration. Simulation results for various log-concave fluxes are shown in Figure~\ref{fig:2}(F).

The difference-measurer strategy is generalized by replacing $\psi_R^{\rm opt}(S)$ with $\psi_R^{\mathrm{opt},\kappa}(S)$ in Eq.~\eqref{eq:general_epsilon}. This class of strategies again converges globally asymptotically stably to $Y^{\rm opt}$, by a similar argument as Theorem 2 . Therefore, our results are robust to the precise kinetics of protein synthesis, and the coarse-grained model captures essential qualitative features.

\clearpage
\onecolumngrid
\section*{Supplemental Material: Optimal Proteome Allocation in a Self-Replicating System By Internal Sensing}

\setcounter{secnumdepth}{3}
\counterwithout*{equation}{section}
\setcounter{section}{0}
\setcounter{equation}{0}
\setcounter{figure}{0}
\setcounter{table}{0}

\renewcommand{\thesection}{S\arabic{section}}
\renewcommand{\thesubsection}{\thesection.\arabic{subsection}}
\renewcommand{\thesubsubsection}{\thesubsection.\arabic{subsubsection}}
\renewcommand{\theequation}{S\arabic{equation}}
\renewcommand{\thefigure}{S\arabic{figure}}
\renewcommand{\thetable}{S\arabic{table}}

\section{Proteome Partition Models}
\subsection{Reaction Network Theory}
Reaction-network theory provides a natural and rigorous language for describing cellular growth by treating
the cell as a collection of interacting biochemical processes.

Based on this, we consider a system connected to an environment $E$, which acts as a reservoir of biomass resources. The system is composed of nodes $ x_1, \dots , x_n $ and \emph{interconversion reactions}, or \emph{fluxes}, $ \phi_0, ..., \phi_m $. Let $ \mathcal{S} := \{1,2,\dots,m\} $ denote the index set of species, and $ \mathcal{R} := \{1,2,\dots,r\} $ denote the index set of reactions. We write $X = (X_1,\dots,X_m) \in \mathbb{R}_{\ge 0}^m$ for the vector of species biomass, where $X_i$ is the biomass of species $i \in \mathcal{S}$.

Each reaction $j \in \mathcal{R}$ is specified by a pair of \emph{complexes} $ a^{(j)} \to a'^{(j)} $, where $a^{(j)}, a'^{(j)} \in \mathbb{R}_{\ge 0}^m$ encode the stoichiometric coefficients of reactants and products, respectively. The \emph{stoichiometric vector} of reaction $j$ is defined by
\begin{equation}
    s^{(j)} := a'^{(j)} - a^{(j)} \in \mathbb{R}^m.    
\end{equation}

The \emph{stoichiometric matrix} $S \in \mathbb{R}^{m \times r}$ can then be written column-wise as $ S = \bigl(s^{(1)} \; s^{(2)} \; \cdots \; s^{(r)}\bigr) $, so that the $j$-th column $S_{\cdot j}$ is precisely the reaction vector $s^{(j)}$. The dynamics is then described by
\begin{equation}
    \label{eq:reaction_network_ode}
    \frac{\mathrm{d} X}{\mathrm{d} t} = S J(X),
\end{equation}
where $ X \in \mathbb{R}^m $ is the vector of species concentrations, and $ J(X) \in \mathbb{R}^r $ is the vector of \emph{reaction fluxes}. In the state space $\mathbb{R}_{\ge 0}^m$, the system of ODEs \ref{eq:reaction_network_ode} defines a vector field $F : \mathbb{R}_{\ge 0}^m \to \mathbb{R}^m$ given by 
\begin{equation}
    F(X) := S J(X), 
\end{equation}
where the components $J_j(X)$ of the flux vector $J(X) = (J_1(X),\dots,J_r(X))^\top$ are nonnegative, i.e. $ J_j : \mathbb{R}_{\ge 0}^m \to \mathbb{R}_{\ge 0} $ whose specific form is determined by the choice of kinetics. In our study, we consider the two most biologically relevant fluxes: Michaelis-Menten type flux and mass action flux. 


\subsection{The Minimal Model}
To capture the autocatalytic nature of a bacterial cell using a minimal model, we need at least two nodes that catalyze each other. Furthermore, we assume that carbon source is the only limiting nutrient.
Consider a model with three nodes $ x_S, x_P, x_R $: a precursor/amino acid node $ x_S $, a transporter-metabolic node $ x_P $, and a ribosomal node $ x_R $ consisting of translation machinery. The respective compositions are $S, P, R$.

We write $ X_{\alpha} $, $\alpha\in\{S,P,R\}$, to denote the biomass in node $ x_{\alpha} $, and $ N = \sum_{\alpha}X_{\alpha}$. The influx and synthesis flux are
\begin{subequations}
    \begin{align}
        J_{\text{in}}(X) &= b X_P, \\
        J_{\text{syn}}(X) &= r X_S X_R / N,
    \end{align}
\end{subequations}
Here, $ r $ is reaction rate, assuming single particle reaction, and $ b $ is the nutrient quality. Typically, $ b $ increases with richer carbon source following Monod's law. The dynamical equations are 
\begin{subequations}
    \begin{align}
        \dot{X}_S &= J_{\text{in}} - J_{\text{syn}}, \\
        \dot{X}_P &= \phi_P J_{\text{syn}}, \\
        \dot{X}_R &= \phi_R J_{\text{syn}},
    \end{align}
    \label{eq:dynamics_minimal}
\end{subequations}
where $ \phi_P, \phi_R $ are the proteome partition fractions for the metabolic and ribosomal sectors, respectively, and $ \mu $ satisfies $ \dot{N}(t) = \mu N(t) $. 
There is a unique stable steady state
\begin{subequations}
    \begin{align}
        S^* &= \left(1 + \frac{r\phi_R}{b\phi_P}\right)^{-1}, \\
        P^* &= \frac{r\phi_R}{b} \left(1 + \frac{r\phi_R}{b\phi_P}\right)^{-1}, \\
        R^* &= \frac{r\phi_R^2}{b\phi_P} \left(1 + \frac{r\phi_R}{b\phi_P}\right)^{-1}.
    \end{align}
\end{subequations}

The steady-state growth rate $\mu^*$ is a harmonic-mean of an uptake-limited rate $ b\phi_P $ and a translation-limited rate $ r\phi_R $:
\begin{equation}
    \mu^* (\phi) = \frac{(b\phi_P)(r\phi_R)}{b\phi_P + r\phi_R}. 
\end{equation}

The optimal fixed strategy is 
\begin{equation}
    \phi_R^{\rm opt} = \frac{\sqrt{b/r}}{1 + \sqrt{b/r}}.
\end{equation}
As $b/a$ increases, the optimal partition shifts towards the ribosomal protein sector, and the cell grows at a rate 
\begin{equation}
    \mu^{\rm opt} = \left(\frac{\sqrt{br}}{\sqrt{b} + \sqrt{r}}\right)^2. 
    \label{eq:lambda_minimal}
\end{equation} 

\subsection{The Two-Sector Model}
Biological processes are in general nonlinear, and we shall later see that introducing saturable nonlinearity tames the deviation from linear growth law. Next, we develop exact theory for the fixed-allocation \emph{two-sector model}. The influx and synthesis fluxes are
\begin{subequations}
    \begin{align}
        J_{\text{in}}(X) &= b X_P, \\
        J_{\text{syn}}(X) &= \left(\frac{a X_S X_R}{k N + X_S}\right),
    \end{align}
\end{subequations}
Here, $ a $ is the maximal translation rate, and $ k $ is the saturation constant. The dynamical equations are identical to those of the minimal model.

In the composition coordinate $Y = (S, P, R)$, 
\begin{subequations}
    \begin{align}
        \frac{\mathrm{d} S}{\mathrm{d} t} &= b P - \frac{a S R}{k + S} - b P S ,\\
        \frac{\mathrm{d} P}{\mathrm{d} t} &= \phi_P \left(\frac{a S R}{k + S}\right) - b P^2 ,\\
        \frac{\mathrm{d} R}{\mathrm{d} t} &= \phi_R \left(\frac{a SR}{k + S}\right) - b P R. 
    \end{align}
\end{subequations}

The growth rate at steady state $\mu^*(\phi_R)$ is given by
\begin{equation}
    \mu^*(\phi_R) = a\phi_R \left( \dfrac{-c(\phi_R)+\sqrt{c(\phi_R)^2 + 4k}}{-c(\phi_R)+2k+\sqrt{c(\phi_R)^2 + 4k}} \right), 
\end{equation}
where $c(\phi_R) \equiv \frac{a\phi_R}{b\phi_P} + k - 1 $. The unique composition steady state $(S^*,P^*,R^*)$ is
\begin{subequations}
    \begin{align}
        S^* &= \frac{1}{2} \left[\sqrt{\left(\frac{a \phi_R}
        {b\phi_P} + k - 1\right)^{2} + 4k} - \left(\frac{a\phi_R}{b\phi_P} + k - 1\right) \right], \\
        P^* &= \left(\frac{a\phi_R}{b}\right) \left(\frac{S}{k + S}\right) \\
        R^* &= \left(\frac{a\phi_R^2}{b\phi_P}\right) \left(\frac{S}{k + S}\right). 
    \end{align}
\end{subequations}

Maximizing $\mu^*(\phi_R)$ with respect to proteome partition $\phi_R$ gives the optimal fixed strategy
\begin{equation}
    \phi_R^{\text{opt}}(\eta) = \frac{\sqrt{\dfrac{kb}{a}}+(1+k)\left(\dfrac{b}{a}\right)}{1+2\sqrt{\dfrac{kb}{a}}+(1+k)\left(\dfrac{b}{a}\right)}.
\end{equation}

Let $ x = \frac{a}{b}\frac{1-\phi_P}{\phi_P}$, $ A = x+k-1 $, $ B = x-k-1 $, and $ R = \sqrt{A^2+4k} $. Then $ \mu^*(\phi) = a (1-\phi_P) (R-A)/(R-B) $. For $\phi_P\in(0,1)$ and $a,k,b>0$, we have $\mu^*>0$, so $\phi_P$ satisfies
\begin{equation}
    \frac{\mathrm{d}}{\mathrm{d}\phi_P}\log \mu^*(\phi_P)=0 
    \implies
    -\frac{1}{1-\phi_P}+\frac{R'-A'}{R-A}-\frac{R'-B'}{R-B} = 0.
    \label{eq:S24}
\end{equation}

Since $A'=B'=x'$ and $ R^{\prime} = A A^{\prime} / R = A x^{\prime} /R $, we obtain
\begin{equation}
    \frac{R'-A'}{R-A}=-\frac{x'}{R}.
    \label{eq:S24e}
\end{equation}
Substitute \eqref{eq:S24e} into \eqref{eq:S24}, note that $B-A=-2k$, we have 
\begin{equation}
    -\frac{1}{1-\phi_P}-\frac{x'}{R}\left(\frac{2k}{R-B}\right) = 0.
    \label{eq:S24g}
\end{equation}
Using $ x'=-\frac{a}{b}\frac{1}{\phi_P^{2}} $, we get
\begin{equation}
    R(R-B) = \frac{2ak}{b}\frac{1-\phi_P}{\phi_P^{2}} = \frac{2kx}{\phi_P}.
    \label{eq:S24k}
\end{equation}
But since $ R^2 - B^2 = 4kx$, we also have 
\begin{equation}
    R(R-B)=\frac{R(R^2-B^2)}{R+B}=\frac{4kx\,R}{R+B},
    \label{eq:S25ter}
\end{equation}
equating the two expressions gives $ R(2\phi_P^{\rm opt}-1) = B $. Substitute $\phi_P=1 /(1+y) $, we have
\begin{equation}
    ay/b-(k+1)=\frac{1-y}{1+y}\sqrt{(ay/b+k-1)^2+4k}
\end{equation}
Squaring gives a quadratic equation in $y$. Pick the physical root $y>0$,
\begin{equation}
    y=\frac{\sqrt{kb/a}+(1+k)b/a}{1+\sqrt{kb/a}},
\end{equation}
then $\phi_R^{\rm opt}=y/(1+y)\in(0,1)$.

More proportion of the proteome is allocated to the $R$-sector as nutrient quality $b/a$ increases, and vice versa. It reduces to the minimal model in the $ k \to \infty $ limit, where the rate constant $ r_{\text{MA}} \equiv \lim_{k \to \infty} \frac{a}{k} $ with $a/k$ kept finite. 

Moreover, the optimal steady-state growth rate is bounded,
\begin{equation}
    \mu^{\rm opt} = \dfrac{b}{1 + 2\sqrt{\dfrac{kb}{a}} + (1+k)\left(\dfrac{b}{a}\right)} \leq \frac{a}{1+k},
\end{equation}
and $\mu^{\rm opt} \to 0 $ linearly as $ b/a $ decreases.

\subsection{Derivations For the Three-Sector Model}
The fixed proteome fractions for each sector $ P, Q, R $ is $ (\phi_P,\phi_Q,\phi_R)\in\Delta_2 $, where $ \phi_Q $ is given. Hence, this model is in fact a modified two-sector model dependent on the optimization of $ (\phi_P, \phi_R)$ subject to $ \phi_P + \phi_R = 1 - \phi_Q$ and $\phi_P,\phi_R >0$. The dynamical equations are
\begin{subequations}
    \begin{align}
        \dot{S} &= b P - \frac{a S R}{k + S} - \mu S, \\
        \dot{P} &= \phi_P \left(\frac{a S R}{k + S}\right) - \mu P, \\
        \dot{Q} &= \phi_Q \left(\frac{a S R}{k + S}\right) - \mu Q, \\
        \dot{R} &= \phi_R \left(\frac{a S R}{k + S}\right) - \mu R,
    \end{align}
\end{subequations}
where $ \mu(t) = b P(t) $ is given by flux balance Eq.~\ref{eq:flux_balance}: Suppose the biomass vector field is $\dot X_{\alpha} = F_{\alpha}(X)$, for $\alpha\in\{S,P,Q,R\}$, then 
\begin{equation}
    \mu = \sum_{\alpha \in \{S, P, Q, R\}} F_\alpha (S, P, Q, R). 
    \label{eq:flux_balance}
\end{equation}

The unique composition steady-state is 
\begin{subequations}
   \begin{align}
       S^* &= \frac{1}{2} \left[ \sqrt{\left(\frac{a \phi_R}{b \phi_P} + k -1\right)^2 + 4k} - \left(\frac{a \phi_R}{b \phi_P}+k-1\right) \right], \\
       P^* &= \left(\frac{a \phi_R}{b}\right) \left(\frac{S}{k + S}\right), \\
       Q^* &= \left(\frac{a \phi_Q \phi_R}{b \phi_P}\right) \left(\frac{S}{k + S}\right), \\
       R^* &= \left(\frac{a \phi_R^2}{b \phi_P}\right) \left(\frac{S}{k + S}\right).
   \end{align}
\end{subequations}

Mke a change of variables $ \phi_P \mapsto\phi_P/q, \phi_R\mapsto \phi_R/q $, where $ q = 1 - \phi_Q $ is the total proteome fraction available for growth-related functions. 
The steady-state growth rate $\mu^*_{\text{canon}} (\phi)$ is related to that of the two-sector model $ \mu^*_{\text{two-sector}} (\phi) $ by
\begin{equation}
    \mu^*_{\text{canon}} (\phi_R) = q \mu^*_{\text{two-sector}} \left(\phi_R/q\right). 
\end{equation}
Therefore, the optimizer of this system should occur at the same $\phi_R/\phi_P$ ratio as the two-sector model, and hence $ \phi^{\text{opt}}_{R,\text{canon}} = q \phi^{\text{opt}}_{R,\text{two-sector}} $, the right hand side of which have already been derived. The optimal fixed strategy is then
\begin{equation}
    \phi_{R}^{\text{opt}} = q \left[\frac{\sqrt{kb/a}+(1+k)b/a}{1 + 2 \sqrt{kb/a} + (1+k)b/a}\right].
\end{equation}
Biologically, $ \phi_Q=1-q $ reduces the total proteome available for growth, and hence the optimal strategy is to allocate the remaining available proteome as if there were only two sectors. The three-sector model reduces to the minimal model as $ k \to \infty $ while keeping $ r \equiv a / k $ finite:
\begin{equation}
    \begin{split}
        \phi_{P, \text{canon}}^{\text{opt}} &= q \left[ \dfrac{1 + \sqrt{kb/a}}{ 1 + 2 \sqrt{kb/a} + (1+k)b/a} \right] \rightarrow q\left[\dfrac{1}{1 + \sqrt{b/r}}\right]. 
    \end{split}
\end{equation}


\subsection{Discussion}
We consider the three-sector model, since there are many proteins belong to housekeeping, stress-response, regulatory, or poorly assigned sectors\cite{Hui2015GlobalResourceAllocation,schmidt_quantitative_2016}, and some proteins are multifunctional or “moonlighting,”\cite{aseev_extraribosomal_2011}. In the three-sector model, ribosome activity\cite{dai_reduction_2016} and degradation\cite{calabrese_protein_2022} are not considered, and the complex biochemical reactions happening within the metabolome are treated as one lumped pool. Adding these into the model would yield more biologically accurate predictions, but would also complicate the dynamical analysis that is central to later sectinos. Finally, the kinetics are assumed to be independent of nutrient abundance, even in conditions where stress response may be induced\cite{valgepea_escherichia_2013}.

\subsection{Parameter Estimate}
We estimate parameter scales for \textit{E. coli} growing near $37^\circ\mathrm{C}$. The $R$-sector consists of ribosomes, RNA, tRNA-charging enzymes (aaRS), translation factors, etc.; The $P$-sector consists of  transporters, catabolic, and anabolic enzymes, etc.
Representative values are
\begin{center}
\begin{tabular}{lll}
\hline
Parameter & Baseline & Exploration range \\
\hline
$a$ & $10\,\mathrm{h}^{-1}$ & $8$--$12\,\mathrm{h}^{-1}$ \\
$b$ & Environment dependent & $0.1$--$10\,\mathrm{h}^{-1}$ \\
$k$ & $0.01$ & $0.003$--$0.03$ \\
$q$ & $0.5$ & $0.4$--$0.6$ \\
\hline
\end{tabular}
\end{center}

\paragraph{Precursor half-saturation $k$.}

Let $[S]=\rho S$, where $S$ is the precursor or amino acid biomass fraction and
$\rho$ is the dry cell density. For bacteria, $\rho\approx 0.3$--$0.5\, \mathrm{g}/\mathrm{mL}$.
A dimensional half-saturation concentration $K_S$ can be converted to composition by 
\begin{equation}
\frac{a[S]}{K_S+[S]}=\frac{aS}{k+S},
\quad k=\frac{K_S}{\rho}.
\end{equation}

It is difficult to measure the coarse-grained quantity $K_S$ directly, but the following estimate may work better: Metabolomics measurements give a total intracellular metabolite concentration of $\approx 300\,\mathrm{mM}$, with amino acids accounting for about $49\%$ of them~\cite{bennett_absolute_2009}.
Using $110\,\mathrm{g/mol}$ per residue gives $c_{\mathrm{AA}}\approx 0.147\,\mathrm{mol/L}$, and hence $\rho_S\approx 0.147(110)\approx 16\,\mathrm{g/L}$. 
Measured cytoplasmic protein concentrations are
$\approx 200$--$320\,\mathrm{g/L}$~\cite{cayley_characterization_1991}. Biomass consists of proteins and the precursor pool, so 
$S\approx\frac{\rho_S}{\rho_{\mathrm{protein}}+\rho_S} \approx 0.05\text{--}0.075.$

Let the translation saturation fraction be $f=\frac{\kappa(S)}{a}=\frac{S}{k+S}$, then $k=S\frac{1-f}{f}$. An illustrative near-saturation range $f=0.7$--$0.9$ gives
$k\approx 0.006\text{--}0.032$. Fast-growth experiments support near-maximal elongation~\cite{dai_reduction_2016}, while it should also be noted that total amino-acid abundance need not be the only variable determining elongation. E.g., a more fine-grained model should consider free amino acids and charged tRNA. This calculation motivates a $k$ of order $10^{-2}$.

Related models provide compatible scales: \cite{Giordano2016} uses $K_R=1\,\mathrm{g/L}$ and an inverse density $1/\rho=0.003\,\mathrm{L/g}$, giving $K_R/\rho=0.003$. \cite{Chure_eLife} reports an effective half-saturation of $80\,\mathrm{mM}$, represented as a protein-normalized mass abundance of $0.03$. These support exploring $k=0.003$--$0.03$.

\paragraph{Translation capacity $a$.}
A ribosome contains approximately $N_{\mathrm{aa},R}=7459$
amino-acid residues in total. At a maximal
elongation speed $v_{\max}\approx 20$--$22\,\mathrm{aa\,s}^{-1}$ (note: for yeasts, it is about $v_{\max}\approx 2$--$5\,\mathrm{aa\,s}^{-1}$),
its protein output per unit ribosomal protein mass is
\begin{equation}
    a=\frac{v_{\max}}{N_{\mathrm{aa},R}}
    \approx 9.65\text{--}10.6\,\mathrm{h}^{-1}.
\end{equation}
These values are consistent with ribosome composition and
experimental elongation measurements~\cite{Chure_eLife,dai_reduction_2016}.
We therefore choose $a=10\,\mathrm{h}^{-1}$ as baseline, and $8$--$12\,\mathrm{h}^{-1}$ for analysis.

\paragraph{Metabolic capacity $b$.}
The parameter $b$ describes net biomass supply per unit metabolic-sector biomass.
For glucose-grown wild-type strains, measured growth rates are approximately $0.5$--$0.7\,\mathrm{h}^{-1}$, glucose uptake rates $j_G = 6.5$--$8.5\,\mathrm{mmol\,g_{DW}^{-1}\,h^{-1}}$, and biomass yields $Y_G = 0.4$--$0.5\,\mathrm{g_{DW}/g_{glucose}}$ ~\cite{fischer_metabolic_2003}.
The uptake and yield measurements give the consistent estimate $\mu^* \approx 0.47\text{--}0.77\,\mathrm{h}^{-1}$. 
For $q=0.5$, $\phi_R=0.1$--$0.2$, and $S^*=0.05$,
the measured growth range gives $P^*\approx 29\%\text{--}38\%$, and $b=\mu^*/P^* \approx 1.3\text{--}2.5\,\mathrm{h}^{-1}$.
For a rich-medium doubling time of $20\,\mathrm{min}$,
$\mu=\ln2/(1/3\,\mathrm{h})\approx 2.08\,\mathrm{h}^{-1}$;
taking $\phi_R=0.25$ with the same $q$ and $S^*$ gives
$b\approx 8.8\,\mathrm{h}^{-1}$. These calculations motivate values of order $1$--$10\,\mathrm{h}^{-1}$ for ordinary lab conditions. 
For comparison, related models use analogous metabolic capacities $1.57$ and $4.76\,\mathrm{h}^{-1}$~\cite{Giordano2016},
or explore $0.01$--$20\,\mathrm{h}^{-1}$~\cite{Chure_eLife}.

\newpage
\section{Derivation of Bacterial Growth Laws}
\subsection{The Minimal Model}
We have 
\begin{equation}
    \phi_R^{\text{opt}} = \frac{\sqrt{b}}{\sqrt{r} + \sqrt{b}} = \sqrt{\frac{\lambda}{r}}.
\end{equation}
This is nonlinear for all $0 < \lambda < r$, so no emergent linearity is present in the minimal model. 

\subsection{The Two-Sector Model}
We derive the growth law for the two-sector model from the fixed allocation perspective. For simplicity of notation, we use $\phi_R$ in place of $\phi_R^{\rm opt}$ where unambiguous. 
Let $ D (x) = 1 + 2 \sqrt{kx} + (1+k)x$. Treat $b$ as a hidden variable, calculate the following quantities:
\begin{equation}
    \phi_R - \frac{\lambda}{a} = \dfrac{\sqrt{kb/a} (1 + \sqrt{kb/a})}{D(b/a)}
    \label{eq:phi_minus_L}
\end{equation}
and
\begin{equation}
    1-\frac{\lambda}{a} = \dfrac{(1+\sqrt{kb/a})^2}{D(b/a)}. 
    \label{eq:one_minus_L}
\end{equation}
Eliminating the dependence on $b$, we get
\begin{equation}
    \left(\phi_R-\frac{\lambda}{b}\right)^2 = \frac{k\lambda}{a}\left(1 - \frac{\lambda}{a}\right).
    \label{eq:phi_L_exact_identity}
\end{equation}
This is a quadratic in $\phi_R$. To choose the physical branch, consider \(\lambda \to 0^+\). Since $\phi_R>0$ aand $\phi_R \simeq \sqrt{k\lambda/a}$ as $\lambda/a\to 0$, $+$ is correct, and 
\begin{equation}
    \phi_R(\lambda/a) = \frac{\lambda}{a} + \sqrt{\frac{k\lambda}{a} \left(1-\frac{\lambda}{a}\right)}.
    \label{eq:phi_lambda_exact_final}
\end{equation}

Now we treat $a$ as the hidden variable. Calculate the following two quantities: 
\begin{equation}
    (1+k)(1-\phi_R) - \frac{\lambda}{b} = \dfrac{\sqrt{ka/b} \left(1 + k + \sqrt{ka/b} \right)}{D(a/b)}, 
\end{equation}
and 
\begin{equation}
    \begin{split}
        (1+k) - \frac{\lambda}{b} = \dfrac{
            \left(1 + k + \sqrt{ka/b}\right)^2
        }{
            D(a/b)
        }. 
    \end{split}
\end{equation}
Eliminating the dependence on $a$, we get
\begin{equation}
    \left[ (1+k)(1-\phi_R)-\frac{\lambda}{b} \right]^2 = k \frac{\lambda}{b} \left(1+k - \frac{\lambda}{b}\right).
\end{equation} 

This is a quadratic in $\phi_R$. To choose the physical branch, consider $a/b\to 0^+$, then $\lambda/b\to 0^+$ while $\phi_R\to 1$. Thus $-$ is correct, and 
\begin{equation}
    \phi_R(\lambda) = 1- \frac{1}{1+k}\left[\frac{\lambda}{b}+\sqrt{\frac{k\lambda}{b}\left(1+k-\frac{\lambda}{b}\right)}\right].
    \label{eq:phi_lambda_dimensional_secondlaw}
\end{equation}

\subsection{The Canonical Model}
\begin{enumerate}
    \item[(1)]
    Convert the two-sector result to the canonical model by $\phi_R \to \phi_R/q$, $\lambda \to \lambda/q$:
    \begin{equation}
        \phi_R = \frac{\lambda}{a} + \sqrt{\frac{k \lambda}{a} \left(q - \frac{\lambda}{a}\right)}, 
        \label{eq:ribosomal_growth_law_HK}
    \end{equation}
    where $0 < \lambda < aq/(1+k) $. Carrying out linear approximation, we have
    \begin{equation}
        y = \frac{1}{a} x + q \left(\frac{\sqrt{k}}{2}\right).
    \end{equation}

    \item[(2)]
    By the same transformation: 
    \begin{equation}
        \phi_R = q - \frac{1}{1+k} \left[\frac{\lambda}{b} + \sqrt{\frac{k\lambda}{b} \left(q (1+k) - \frac{\lambda}{b}\right)} \right], 
        \label{eq:second_law_housekeeping_dimensional}
    \end{equation}
    where $ 0 < \lambda < bq $. Carrying out linear approximation, we have
    \begin{equation}
        y = q\left(1 - \frac{\sqrt{k}}{2}\right) - \frac{1}{(1+k)b} x.
    \end{equation}
\end{enumerate}

\subsection{Temperature-Independence of Growth Laws}
Assume that translation rate and nutrient abundance depend on temperature in a similar way, i.e. $a(T) = a_0 \zeta(T)$, $b(T) = b_0 \zeta(T)$. For example, $\zeta(T) = e^{-E_{\rm act}/(k_BT)}$ may be of the Arrehius form. Then, all quantities involing $b/a$ is temperature-independent, in particular $\lambda(T) = \lambda_0 \zeta(T)$. Consequently, $\lambda/a$ and $\lambda/b$ are temperature-independent, and hence also $\phi_R$. Therefore, growth law is predicted to be temperature independent, consistent with \cite{mairet_optimal_2021}. 

\subsection{Growth Law Trend}

\begin{figure}[htbp]
    \centering
    \includegraphics[width=0.9\textwidth]{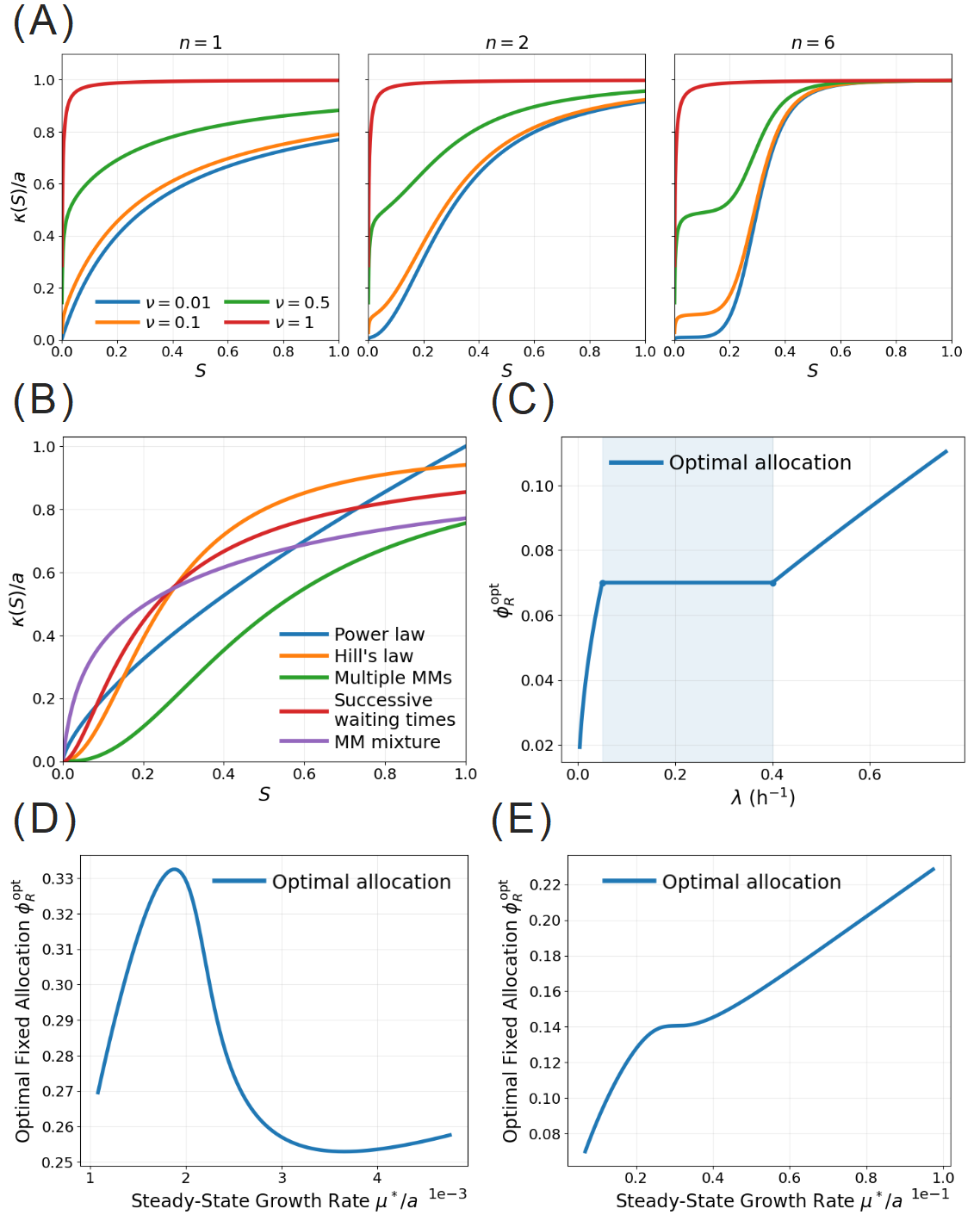}
    \caption{(A) Activated translation kinetics as described by Eq.~\eqref{eq:activated}, simulated for $n=1,2,6$ and activation values $\nu$ from $0.01$ to $1.0$. (B) Log-concave biosynthetic kinetics as listed in Table~\ref{tab:log_concave_translation}. With the exception of the power law, all kinetics exhibit saturation for large $S$. (C) Piecewise $\kappa(S)$ that gives rise to a growth law with a plateau at $\phi_R^{\rm opt}=0.07$, for $\lambda$ in $0.05$--$0.4\,\mathrm{h}^{-1}$. $a$ is chosen to be $10\,\mathrm{h}^{-1}$. (D) Growth law from Eq.~\eqref{eq:activated} with $a=10\,\mathrm{h}^{-1}$, $b = 0.003$--$0.030\,\mathrm{h}^{-1}$, $k=2.85$, $K=1$, activation fraction $\nu=0.4$, and Hill coefficient $n=8$. High degree of nonlinearity gives rise to a highly nonmonotonic growth law. (E) Growth law from Eq.~\eqref{eq:activated} with $a=10\,\mathrm{h}^{-1}$, $b = 0.01$--$0.25\,\mathrm{h}^{-1}$, $k=0.3$, $K=0.28$, activation fraction $\nu=0.5$, and Hill coefficient $n=5$. The growth law exhibit small window of plateau near $\lambda=0.2$--$0.4\,\mathrm{h}^{-1}$.}
    \label{fig:kinetics}
\end{figure}

Reaction fluxes are observed to be \emph{concave}~\cite{yamagishi_global_2025}, here the statement is generalized the weaker \emph{log-concave} condition. Commong biological fluxes that obey \emph{log-concavity} include (1) power-law kinetics; (2) Hill's law; (3) multiple saturating factors; (4) successive waiting times; and (5) sum of Michaelis-Menten reactions. They are listed in Table~\ref{tab:log_concave_translation}, and plotted in Figure~\ref{fig:kinetics}(B).

\begin{table}[htbp]
    \centering
    \renewcommand{\arraystretch}{2}
    \begin{tabular*}{\textwidth}{@{\extracolsep{\fill}}lcc@{}}
        \toprule
        Kinetics (with references) & Specific flux $\kappa(S)$ & Conditions\\
        \midrule
        Power law
        & $\displaystyle a S^n $ & $n>0$
        \\[3pt]
        Hill saturation
        & $\displaystyle
            a\frac{S^n}{K^n+S^n}$ & $n>0$
        \\[3pt]
        Multiple saturation factors
        & $\displaystyle
            a\prod_j\frac{S^{n_j}}{K_j^{n_j}+S^{n_j}}$
        & $n_j>0$
                \\[3pt]
        Successive waiting times
        & $\displaystyle
            \left(\tau_0+\sum_j c_jS^{-p_j}\right)^{-1}$
        & $\tau_0,c_j,p_j>0$
        \\[3pt]
        Michaelis-Menten mixture~\cite{Weiss2015Mechanistic,yamagishi_global_2025}
        & $\displaystyle
            \sum_j \frac{a_j S}{K_j+S}$
        & $a_j,K_j>0$
        \\
        \bottomrule
    \end{tabular*}
    \caption{Examples of log-concave translation-rate functions. Here $S>0$ is the metabolite fraction, and all rate amplitudes and saturation constants are positive.}
    \label{tab:log_concave_translation}
\end{table}

We modify the biosynthetic kinetics with a small and sufficiently nonlinear term. E.g.,
\begin{equation}
    \kappa(S) = \left(\frac{aS}{k+S}\right)
    \left[\nu + (1-\nu)\left(\frac{S^n}{K^n + S^n}\right)\right],
    \label{eq:activated}
\end{equation}
where $0<\nu<1$ represents the activation strength and $n\in\mathbb{N}$. The function is plotted in Figure~\ref{fig:kinetics}(A). In the minimal $n=2$ case, we have $\kappa(S) = aS(S^2+\nu K^2)/[(k+S)(K^2+S^2)]$. The metabolite composition $S^{\rm opt}$ that maximizes $\mu^*(S^{\rm opt})$ is the root of the sextic
\begin{equation}
    ax^2 (\nu K^2 + x^2)^2 = b(1-x^2)
    \left[kx^4 + 2(1-\nu)K^2 x^3 + (3-\nu)kK^2x^2 + \nu kK^4\right].
\end{equation}
$\phi_R^{\rm opt}$ and $\mu^{\rm opt}$ can be computed numerically. We find a (1) flat and (2) negatively correlated quasi-linear growth law within a small range.

We can construct a piecewise strategy that optimizes steady-state growth rate with growth-independent allocation for moderate growth rate. From the precursor formulation, 
\begin{equation}
    \phi_R^{\rm opt}(S) = q\left[\frac{1}{1+(1-S)(\log\kappa(S))'}\right].
\end{equation}
Suppose the strategy is fixed at $\phi_0\in(0,1)$, we pick $p=q/\phi_0-1$ and $\kappa(S)=C(1-S)^{-p}$, where $C>0$ is a constant. An example simulation is shown in Figure~\ref{fig:kinetics}(C), where $\phi_0$ is set to $0.05$, corresponding to an RNA/protein mass ratio of about $0.07/0.46 = 0.15$ for \textit{E. coli}~\cite{Scott2010}. This reproduces the experimental measurements in \cite{matamouros_growth_rate_2023}, where the growth law curve stays constant at RNA/protein$\approx 0.15$ for growth rates under $0.4\,\mathrm{hr}^{-1}$. A globally growth-independent strategy is impossible since $\kappa(0)=0$. In Figures~\ref{fig:kinetics}(D)(E), we demonstrate two cases of Eq.~\eqref{eq:activated} with a nonmonotonic and plateauing growth law, respectively.

\newpage
\section{Dynamical Analysis}
\subsection{Fixed and Adaptive}
We demonstrate the correspondence between the fixed and the adaptive system using the proteome partition model. All discussion here can be generalized to other reaction networks. 

The adaptive system with strategy $\phi_R(S,P,Q,R)$
\begin{subequations}
    \begin{align}
        \frac{\mathrm{d}S}{\mathrm{d}t} &= bP - \frac{aSR}{k+S} - bPS, \\
        \frac{\mathrm{d}P}{\mathrm{d}t} &= \phi_P (S,P,Q,R) \left(\frac{aSR}{k+S}\right) - bP^2, \\
        \frac{\mathrm{d}Q}{\mathrm{d}t} &= \phi_Q \left(\frac{aSR}{k+S}\right) - bPQ, \\
        \frac{\mathrm{d}R}{\mathrm{d}t} &= \phi_R (S,P,Q,R) \left(\frac{aSR}{k+S}\right) - bPR, 
    \end{align}
\end{subequations}
if converges, converges to a steady state that is in general different from that of the original constant-partition model. Given a convergent system with $(S,P,Q,R) \to (S^*,P^*,Q^*,R^*)$ and strategy $\phi(S,P,Q,R)$, let the fixed strategy be 
\begin{equation}
    \phi_R \equiv \phi_R(S^*,P^*,Q^*,R^*).
\end{equation}
The two systems converge to the same linear reaction network (LRN). The analysis until now also applies to suboptimal steady states. 

Dynamics of the adaptive system under $\phi_R (Y)$ can be written as $\dot{X}=M(Y)X$, where
\begin{equation}
    M(Y)=
    \begin{pmatrix}
    0 & b & 0 & -\kappa(S)\\
    0 & 0 & 0 & \phi_P(Y)\kappa(S)\\
    0 & 0 & 0 & \phi_Q\kappa(S)\\
    0 & 0 & 0 & \phi_R(Y)\kappa(S)
    \end{pmatrix}.
    \label{eq:adaptive_matrix}
\end{equation}

The system is nonlinear because the coefficients of $M$ depend on $Y$. Suppose the composition converges to an interior balanced-growth steady state, $Y(t)\to Y^*$. By continuity,
\begin{equation}
    M(Y(t))\longrightarrow M(Y^*)\equiv M^*,
\end{equation}
so the adaptive system asymptotically reduces to a constant-coefficient LRN. Similarly, $\phi_R(Y)\to\phi_R(Y^*) \equiv \phi_R$, which defines the corresponding fixed-allocation model. Conversely, a fixed steady state can be realized by an adaptive system only when $\phi_R(Y^*)=\phi$.

\subsection{Suboptimal Allocation With Two Parameters}
At any balanced-growth steady state, define the ratio of ribosomal to transporter-and-metabolic protein expression at steady state as 
\begin{equation}
    \omega^*=\frac{R^*}{P^*}, 
    \qquad 
    \phi_R=\frac{q\omega^*}{1+\omega^*}.
\label{eq:omega_relation}
\end{equation}
Mtabolite balance further gives
\begin{equation}
    \frac{b}{a}=\frac{S^*\omega^*}{(k+S^*)(1-S^*)}.
    \label{eq:steady_balance}
\end{equation}
For a general suboptimal balanced state, $S^*$ alone is insufficient without knowledge of the environment $\eta$. Indeed, Eq.~\eqref{eq:steady_balance} shows that the same $S^*$ may be associated with different $\omega^*$, and hence with different allocations. On the other hand, the pair $(S^*, \omega^*)$ is sufficient. 

\begin{figure}[htbp]
    \centering
    \includegraphics[width=0.8\textwidth]{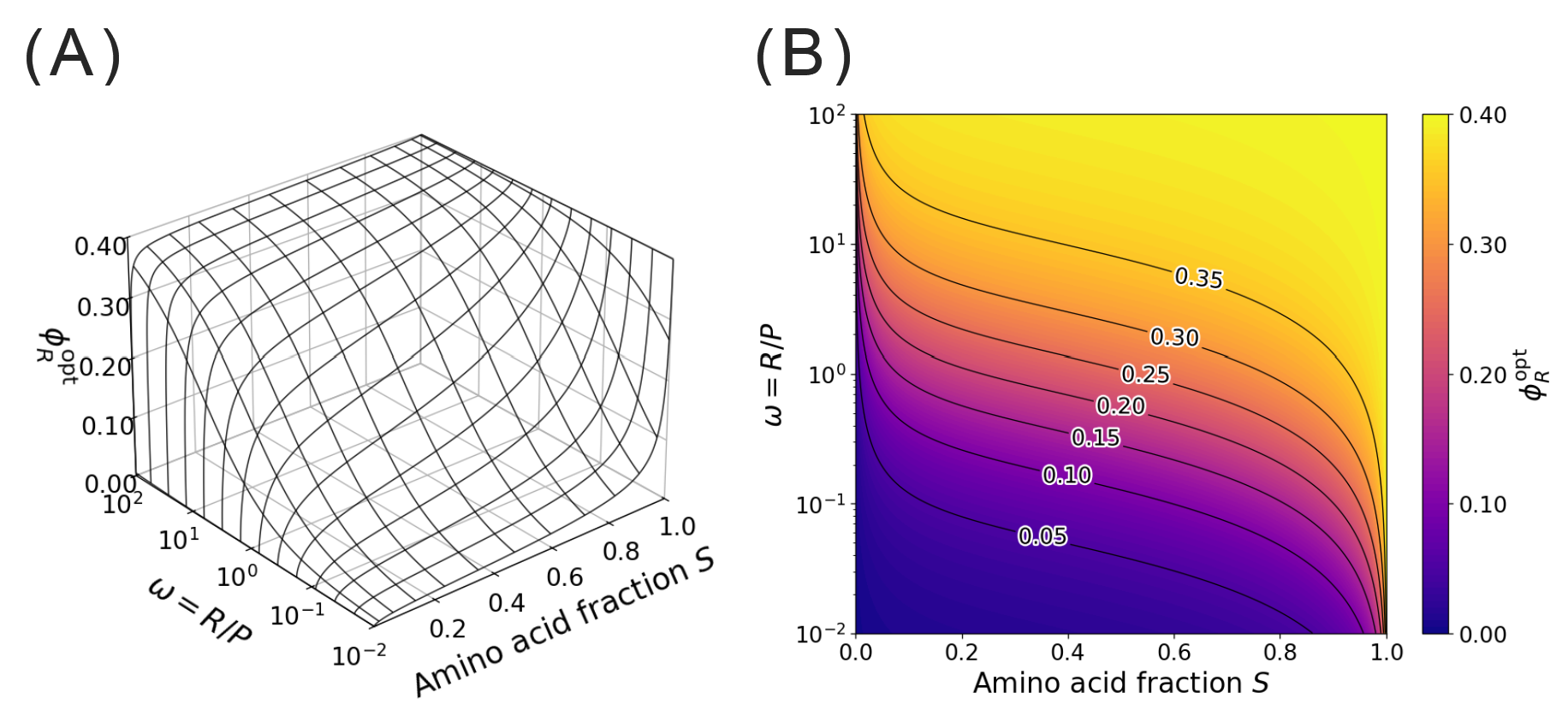}
    \caption{(A) 3D contour of the optimal fixed strategy $\phi_R^{\rm opt}(S^*,\omega^*)$. The precursor composition ranges from $0$ to $1$, while expression ratio ranges from $10^{-2}$ to $10^2$ on a logarithmic scale. (B) Contour diagram of (A).}
    \label{fig:two_var}
\end{figure}

Given a possibly suboptimal steady state $(S^*, \omega^*)$, the optimal fixed strategy is 
\begin{equation}
    \phi_R^{\rm opt}(S^*,\omega^*) = q \left[\dfrac{\sqrt{\dfrac{kS^* \omega^*}{(k+S^*)(1-S^*)}} + \dfrac{(1+k)S^* \omega^*}{(k+S^*)(1-S^*)}}{1 + 2\sqrt{\dfrac{kS^* \omega^*}{(k+S^*)(1-S^*)}}+ \dfrac{(1+k)S^* \omega^*}{(k+S^*)(1-S^*)}}\right].
    \label{eqs:optimal_two_param}
\end{equation}
The dependence is plotted in Figure~\ref{fig:two_var} for $q=0.4$ and $k=0.11$.  

\subsection{Adaptive Optimization}

Substituting $b/a = S^2/[k(1-S)^2]$ into the optimal fixed strategy, we eliminate $b/a$ and find an single-signal optimal adaptive strategy 
\begin{equation}
    \psi_R^{\rm opt}(S)=q\left[\frac{S(k+S)}{k+S^2}\right].
    \label{eq:optimal_sensing}
\end{equation}

This amino-acid-sensing strategy converges to a unique stable optimal steady state, by applying Theorem 2 of the main text. Optimal allocation can be studied from the perspective of a dynamical variable. An insightful choice is amino acid composition $S^*$. 

\begin{proposition}
    When the environment $\eta = \{A,b\}$ is known, where $b$ is nutrient quality and $A$ is a collection of variables associated with the specific translation rate $\kappa(S)$, subject to some biological constraints, optimal allocation can be decided. 
\end{proposition}
\begin{proof}
    The $S^*$-controlled steady-state growth rate $\mu^*(\eta)$ is
    \begin{equation}
        \mu^*(S^*) = q \left[\dfrac{b(1-S^*)\kappa(S^*)}{b(1-S^*) + \kappa(S^*)}\right] = q \left[\frac{1}{b(1-S^*)} + \frac{1}{\kappa(S^*)}\right]^{-1},
    \end{equation} 
    where the steady state $S^*$ is to be interpreted as a number. Minimizing the bracket gives  
    \begin{equation}
        \label{eq:growth_rate_S}
        \frac{\mathrm{d}(q/\mu^*)}{\mathrm{d}S^*} = \frac{1}{b(1-S^*)^2} - \frac{\kappa'(S^*)}{\kappa(S^*)} = 0. 
    \end{equation}

    If the cell regulates $\mu^*$ directly, the strategy solves a scalar optimization problem in $S^*$. 
    
    When $\kappa(S) = aS/(k+S)$, $S^{\rm opt} = \sqrt{kb/a} / \left(1+\sqrt{kb/a}\right)$. This then gives an alternate derivation of $\phi_R^{\rm opt}$ in terms of $\eta$, as shown before by a tedious computation.
\end{proof}

\begin{proposition}[balanced growth rate upper bound]
The optimal constant-allocation growth rate, under the assumption that $\eta$ is known, 
\begin{equation}
    \mu^{\rm opt}=\frac{q b}{1+2\sqrt{kb/a}+(1+k)b/a}
\end{equation}
is an upper bound on the growth rate of any converged interior adaptive steady state. This bound does not apply to instantaneous transient growth rates.
\end{proposition}

\begin{proof}
    Since $\mu^{\rm opt}$ is a global optimum of the constant system over all fixed strategies, and the adaptive system -- if converges -- converges to the corresponding fixed system, the proposition is true. Transiently, $\mu(t) = bP(t) \le b$ is also globally bounded, but $\mu(t)$ need not be bounded by $\mu^{\rm opt}<b$. 
\end{proof}

\subsection{Proof of Main Theorems}

\begin{proposition}[forward invariance and positivity]
    In proteome partition, the space in which the composition $(S,P,Q,R)$ lies in is forward invariant and $S,P,Q,R>0$ for all $t$.
\end{proposition}
\begin{proof}
    Summing the dynamical equations gives $\frac{\mathrm{d}}{\mathrm{d}t}(S+P+Q+R)=bP(1-S-P-Q-R)$, so initial normalization implies $S+P+Q+R=1$ for all $t>0$. On each boundary face, the derivative of the component is nonnegative, so $\Delta_3$ is forward invariant. Moreover,
    \begin{equation}
        \dot S\ge-\frac{a}{k}S,
        \qquad
        \dot P\ge-bP,
        \qquad
        \dot Q\ge-bQ,
        \qquad
        \dot R\ge-bR.
    \end{equation}
    on $\Delta_3$. Hence, every initially positive component therefore remains positive at finite times. The vector field is smooth on the compact simplex, so solutions exist globally.
\end{proof}

Since $(S,P,Q,R)$ is normalized, there are only three degrees of freedom. Define
\begin{equation}
    v=\frac{qR}{P+R} \in (0,q),
    \qquad 
    w = \frac{P+R}{1-S} \in (0,1),
\end{equation}
with inverse transformation
\begin{equation}
    P=(1-S)w\left(1-\frac{v}{q}\right),\qquad
    Q=(1-S)(1-w),\qquad
    R=(1-S)\frac{wv}{q}.
    \label{eq:PQR}
\end{equation}

Define $G(S)=(1-S) [b(1-S)+\kappa(S)]$ and $r_0(S)=\frac{qb(1-S)}{b(1-S)+\kappa(S)}$. The reduced system
\begin{subequations}
    \begin{align}
        \dot S &= \frac{w}{q}G(S)[r_0(S)-v],
        \label{eq:S}\\
        \dot v &= \kappa(S)v(\phi_R-v),
        \label{eq:v}\\
        \dot w &= \frac{\kappa(S)wv}{q}(q-w).
        \label{eq:w}
    \end{align}
\end{subequations}
    
Furthermore,
\begin{equation}
    r_0'(S) = -\frac{qb[\kappa(S)+(1-S)\kappa'(S)]}{[b(1-S)+\kappa(S)]^2}<0,
\end{equation}
with $r_0(0)=q$ and $r_0(1)=0$. Hence $r_0$ is invertible on $(0,q)$ with $r_0^{-1}:(0,q)\to(0,1)$.

\begin{theorem}
    The canonical proteome partition model is persistent. 
\end{theorem}
\begin{proof}
    Eq.~\eqref{eq:v} implies that $v$ moves toward $\phi_R$
    monotonically without crossing it. So,
    \begin{equation}
        0<v_-\equiv\min\{v(0),\phi_R\}
        \le v(t)\le
        v_+\equiv\max\{v(0),\phi_R\}<q.
    \end{equation}
    Similarly, Eq.~\eqref{eq:w} gives
    \begin{equation}
        0<w_-\equiv\min\{w(0),q\}
        \le w(t)\le
        w_+\equiv\max\{w(0),q\}<1.    
    \end{equation}
    These bounds do not assume convergence of $v$ or $w$, or the persistence of $(S,v,w)$. Let 
    \[
    S_-=\min\{S(0),r_0^{-1}(v_+)\}>0,
    \qquad
    S_+=\max\{S(0),r_0^{-1}(v_-)\}<1.
    \]
    Monotonicity of $r_0$ on $[S_-,S_+]$ gives $r_0(S_-)-v(t)\ge v_+-v(t)\ge0$ and $r_0(S_+)-v(t)\le v_--v(t)\le0$.
    Equation~\eqref{eq:S} therefore prevents $S$ from its bounds: $0<S_-\le S(t)\le S_+<1$. Hence, $(S,v,w)$ is persistent, and by Eq.~\eqref{eq:PQR}, so is $(S,P,Q,R)$. Each interior trajectory $Y(t)\subseteq\Delta_3^{\circ}$ remains in a compact subset
    of $\Delta_3^{\circ}$.
\end{proof}

\begin{theorem}[convergence]
    Let $v=qR/(P+R)$, $w=(P+R)/(1-S)$, and let $\phi_R(S,v,w)$ be a (not necessarily optimal) strategy. If the trajectories $Y(t)$ are persistent, and $F(S,v) = \phi_R|_{w=q}$ satisfies $\partial_S F \ge 0$ and $\partial_v F < 1$, then the system converges to the unique interior steady state for every $Y(0)\in\Delta_3^{\circ}$.
\end{theorem}

\begin{proof}
Persistence implies $\kappa(S)wv/q$ is bounded below by a positive constant. Hence
\begin{equation}
w(t)-q = [w(0)-q]\exp\left[-\int_0^t\mathrm{d}\tau\, \frac{\kappa(S(\tau))w(\tau)v(\tau)}{q}\right] \to 0.
\end{equation}
The $\omega$-limit set of each $Y(t)\subseteq\Delta_3^{\circ}$ is therefore nonempty, compact, and contained in $w=q$.

On this invariant manifold, the system reduces to
\begin{subequations}
    \begin{align}
        \dot S &= G(S)[r_0(S)-v], \\
        \dot v &= \kappa(S)v[F(S,v)-v].
    \end{align}
\end{subequations}

There is at most one interior equilibrium, since $S^*$ is a root of 
\begin{equation}
    H(S)\equiv F(S,r_0(S))-r_0(S),
\end{equation}
and both $r_0$ and $H$ are monotonic. In particular, $r_0'(S)<0$,
\begin{equation}
    H'(S)=\partial_S F(S,r_0(S))+[1-\partial_v F(S,r_0(S))][-r_0'(S)]>0.
\end{equation}

Let $B(S,v)=\frac{1}{G(S)v}$ be the Dulac multiplier. Then
\begin{equation}
    \nabla\cdot\left(B(\dot S,\dot v)\right) = \frac{\partial}{\partial S}(B\dot S)
+
\frac{\partial}{\partial v}(B\dot v)
=
\frac{r_0'(S)}{v}
+
\frac{\kappa(S)}{G(S)}[F_v(S,v)-1]
<0.
\end{equation}

The Bendixson--Dulac criterion excludes periodic orbits
in the interior. 

Choose \(z\in\Omega\). Its forward orbit follows the planar system and remains in the compact interior set \(\Omega\). Its \(\omega\)-limit set \(\omega(z)\subseteq\Omega\) must contain an equilibrium, since otherwise the Poincaré–Bendixson theorem would make \(\omega(z)\) a periodic orbit, contrary to the Dulac criterion. Thus \(\Omega\) contains an interior equilibrium, which is unique since \(H'>0\). Denote the unique equilibrium by $(S^*,v^*, w^*)$, where $v^*=r_0(S^*)$ and $w^*=q$.

The planar Jacobian 
\begin{equation}
    J^*=
    \begin{pmatrix}
    G r_0' & -G\\
    \kappa v^* \partial_S F & \kappa v^*(\partial_v F-1)
    \end{pmatrix}_{(S^*,v^*)}
\end{equation}
at $(S^*,v^*)$ satisfies $\operatorname{tr}J^*<0$ and $\det J^*=
G(S^*)\kappa(S^*)v^* H'(S^*)>0$. So it is (locally) asymptotically stable. 

The full Jacobian is given by 
\begin{equation}
    J_{\mathrm{full}}^*=
    \begin{pmatrix}
    Gr_0'&-G&0\\
    \kappa v^*F_S&\kappa v^*(F_v-1)&
    \kappa v^*\partial_w\phi_R\\
    0&0&-\kappa v^*
    \end{pmatrix}_{(S^*,v^*,q)}.
\end{equation}
\end{proof}

\begin{theorem}[difference-measurer strategy]
    Let $\varepsilon\in C^{1}((0,q),\mathbb{R})$ satisfy $0<\varepsilon'(\cdot)\le 1$, and let $\psi(S)=q\frac{S(k+S)}{k + S^2}$. Then the strategy
    \begin{equation}
        \label{eq:general_epsilon}
        \phi_{R}^{\rm opt, [\varepsilon]} (Y) = \frac{qR}{P+R} + \varepsilon\left(\psi(S)\right) 
        - 
        \varepsilon\left(\frac{qR}{P+R}\right)
    \end{equation}
    is optimal. We call it the \emph{difference-measurer}.
\end{theorem}

\begin{proof}
\noindent \textbf{(1) Admissibility:} By the Mean Value Theorem, when $\psi(S)\neq v$, 
\begin{equation}
    \phi_{R}^{\rm opt, [\varepsilon]} = v + \varepsilon\left(\psi(S)\right) - \varepsilon\left(v\right)
    = v + \theta\left[\psi(S) - v\right]
\end{equation}
with $0<\theta=\varepsilon'(c)\le 1$. Then, $\phi_R = (1-\theta)v+\theta\psi(S)\in(0,q)$. 

\noindent \textbf{(2) Persistence:}

Unique fixed point: At an interior equilibrium, Eq.~\eqref{eq:w} gives
$w=q$. Strict monotonicity of $\varepsilon$ implies $\dot v = 0$ if and only is $v=\psi(S)$, and Eq.~\eqref{eq:S} requires $v=r_0(S)$. Since $\psi: 0\uparrow q$ and $r_0: q\downarrow 0$, they intersect uniquely at $S^*=\frac{\sqrt{kb/a}}{1+\sqrt{kb/a}}$ and $v^*=\psi(S^*)=r_0(S^*)$. This maximizes
\begin{equation}
    \mu^*(S) = \dfrac{q}{\left(b(1-S)\right)^{-1}+\kappa(S)^{-1}},
\end{equation}
and thus $(S^*,v^*,w^*)$ exists uniquely and is optimal. 

Now define the energy functions
\begin{subequations}
    \begin{align}
        E_S(S) &= \int_{S^*}^{S} \mathrm{d}s\, \frac{\kappa(s)}{G(s)} \bigl[\varepsilon(\psi(s))-\varepsilon(v^*)\bigr] \ge 0, \\
        E_v(v) &= v-v^*-v^*\log\left(\frac{v}{v^*}\right) \ge 0, \\
        W &= \frac{q}{w}E_S(S)+E_v(v).
    \end{align}
\end{subequations}
Both $E_S$ and $E_v$ vanish only at $S=S^*$ and $v=v^*$, respectively. Let $W_S=\frac{q}{w}E_S(S)$,
\begin{subequations}
    \begin{align}
        \dot W_S &= \kappa(S)[\varepsilon(\psi(S))-\varepsilon(v^*)][r_0(S)-v]-\frac{\dot w}{w}W_S,\\
        \dot E_v &= \kappa(S)(v-v^*) [\varepsilon(\psi(S))-\varepsilon(v)].
    \end{align}
\end{subequations}
Consequently,
\begin{equation}
    \begin{split}
        \dot W &= -\kappa(S)\left\{[\varepsilon(\psi(S))-\varepsilon(v^*)] [v^*-r_0(S)]+(v-v^*)[\varepsilon(v)-\varepsilon(v^*)]\right\} -\frac{\dot w}{w}W_S \\
        &\equiv - \kappa(S)D(S,v) - \frac{\dot w}{w}W_S.
    \end{split}
\end{equation}
Both products in $D(S,v)$ are nonnegative: For the first, both factors have the sign of \(S-S^*\); for the second, both have the sign of \(v-v^*\). So $D>0$.

Eq.~\eqref{eq:w} implies $w$ moves monotonically toward $q$ and remains in $[w_-,w_+] \equiv \left[\min\{w(0),q\}, \max\{w(0),q\}\right]$. There are two cases: (1) $w(0)\le q$ and (2) $w(0)>q$.
\begin{enumerate}
    \item If $w(0)\le q$, then $\dot w \ge 0$, and $\dot W \le 0$. Hence, $W(t)\le W(0)$.
    \item If $w(0)>q$, then $\dot w \le 0$ and $w(t)\ge q>0$. Then
    \begin{equation}
        \begin{split}
            \frac{\mathrm{d}}{\mathrm{d}t}(wW) &= \dot w W + w\dot W \\
            &= \dot w (W_S+E_v) + w\left[- \kappa(S)D(S,v) - \frac{\dot w}{w}W_S\right] \\
            &= -w\kappa(S)D(S,v) + \dot w E_v 
            \le 0.
        \end{split}
    \end{equation}
    Then, $W(t)\leq\frac{w(0)}{w(t)}W(0)\leq\frac{w(0)}{q}W(0)$.
\end{enumerate}
Choose $M \equiv \max\left\{1, \frac{w(0)}{q}\right\}W(0)$, we have $W(t)\le M$.

Use $M$ to find $S_+$ and $v_-$: Note that $E_S\to\infty$ as $S\to 1^-$ and $E_v\to\infty$ as $v\to 0^+$. Because \(0<\varepsilon'\leq1\), \(\varepsilon\) extends continuously to \([0,q]\) and $A\equiv \varepsilon(q)-\varepsilon(v^*)>0$. To see the first property, notice that as $S\to 1^-$,
\begin{equation}
    \frac{\kappa(S)}{G(S)} = \frac{\kappa(S)}{(1-S)[b(1-S)+\kappa(S)]} \simeq \frac{A}{1-S}.
\end{equation}
However, $E_S\le \frac{w_+}{q}M$ and $E_v\le W \le M$, so there exist $S_+<1$ and $v_->0$ such that $S\le S_+$ and $v\ge v_-$.

Let $v_+=\max\{v(0),\psi(S_+)\}$. From Eqs.~\eqref{eq:S} and \eqref{eq:v}, whenever $v\geq v_+$, we have
\begin{equation}
    \psi(S)\leq\psi(S_+)\leq v,    
\end{equation}
and hence $\dot v\leq0$. Then, $v(t)\leq v_+<q$. Next, let $S_-=\min\{S(0), r_0^{-1}(v_+)\}>0$. At $S=S_-$, we have $r_0(S_-) \ge v_+ \ge v(t)$, Eq.~\eqref{eq:S} tells us that $\dot S\ge 0$, and thus $S(t)\ge S_- >0$. Finally,
\begin{equation}
    0<S_-\leq S\leq S_+<1,\quad
    0<v_-\leq v\leq v_+<q,\quad
    0<w_-\leq w\leq w_+<1
\end{equation}
implies persistence. 

\noindent \textbf{(3) Partial derivative conditions:} On $w=q$, $F(S,v)=v+\varepsilon(\psi(S))-\varepsilon(v)$. Then, 
\begin{subequations}
    \begin{align}
        \partial_S F &= \varepsilon'(\psi(S))\psi'(S)>0,\\
        \partial_v F &=1-\varepsilon'(v)<1.
    \end{align}
\end{subequations}
Theorem 1 then implies convergence.
\end{proof}

\subsection{Limit Cycles Under Strong Feedback}

We give an admissible strategy that generates a stable limit cycle. Choose an arbitrary $S_c\in(0,1)$ and $R,\gamma>0$. Let $\sigma \equiv S-S_c$ and $\rho\equiv r-r_0(S)$ be the deviations of metabolite and normalized ribosome composition, respectively. Consider
\begin{equation}
    \phi_R(S,r)=r+\frac{A(S)}{\kappa(S)r}\left[\sigma -r_0'(S)\rho +\gamma\rho\left[R^2-\sigma^2-\rho^2\right]\right].
    \label{eq:limit_cycle_policy}
\end{equation}

For sufficiently small $R$, the resulting orbit remains in the interior of the state space and satisfies
$0<\phi_R<q$. The  quantity $V(S,r)\equiv\sigma^2+\rho^2$ satisfies $\dot V=2\gamma A(S) \rho^2\left(R^2-V\right)$. Since $A(S)>0$ in the interior, trajectories with $V<R^2$ are driven outward, whereas trajectories with $V>R^2$ are driven inward. Therefore, $\sigma^2+\rho^2=R^2$ is an invariant attracting closed orbit, on which the dynamics reduce to
\begin{subequations}
    \begin{align}
        \dot S &= -A(S)\rho(r,S),\\
        \dot{\rho} &= A(S)\sigma(S),
    \end{align}
\end{subequations}
so the vector field is nonzero everywhere, and the curve is a periodic orbit.

Biologically, $\rho(r,S)$ measures the imbalance between translational capacity and metabolic supply. The strong-feedback strategy is an example where sufficiently strong state-dependent feedback overcorrects imbalance and produces a stable oscillation.

\newpage
\section{Limited Information in a Multiresource Network}
For the proteome partition model, a single-signal optimal strategy exists. This is not always the case: Consider the complimentary resource model of a cell growing on carbon- and nitrogen-based nutrients.

At balanced growth, $J_{\rm syn}=\mu (1 - S_C - S_N)$. Then, the two precursor equations give $b_CP_C=\mu(\nu_C (1 - S_C - S_N)+S_C)$ and $b_NP_N=\mu(\nu_N (1 - S_C - S_N)+S_N)$. Hence,
\begin{equation}
    \begin{split}
        \frac{1}{\mu} = \frac{1}{b_C}\left(\nu_C+\frac{S_C}{1 - S_C - S_N}\right)
        +\frac{1}{b_N}\left(\nu_N+\frac{S_N}{1 - S_C - S_N} \right)+\frac{R}{J_{\text{syn}}(S_C,S_N)}.
    \end{split}
\end{equation}

\begin{proposition}[control-sufficient signals]
    Sensing $(S_C,S_N)$ is sufficient to reconstruct the optimal allocation, but $(S_C,P_C)$ is not sufficient for doing so.
\end{proposition}
\begin{proof}
Minimizing reciprocal growth rate requires $\partial_{S_C} (1/\mu) = \partial_{S_N} (1/\mu) = 0$. Then, 
\begin{equation}
    \begin{pmatrix}
        1-S_N & S_N \\
        S_C & 1-S_C
    \end{pmatrix}
    \begin{pmatrix}
        1/b_C \\ 1/b_N
    \end{pmatrix} = 
    -(1-S_C-S_N)^2 
    \begin{pmatrix}
        \partial_{S_C} g \\ \partial_{S_N} g
    \end{pmatrix}.
\end{equation}
The determinant of the matrix on the left-hand side is $1-S_C-S_N$. Explicitly inverting:
\begin{subequations}
    \begin{align}
        \frac{1}{b_C} &= -(1 - S_C - S_N) \left[(1-S_C)g_{S_C}-S_N g_{S_N}\right], \\
        \frac{1}{b_N} &= -(1 - S_C - S_N) \left[(1-S_N) g_{S_N}-S_C g_{S_C}\right].
    \end{align}
\end{subequations}
Thus, $(S_C,S_N)$ is a sufficient set of internal signals.
On the other hand, $(S_C,P_C)$ is not sufficient, as the numerical example in the main text explains.
\end{proof}

\begin{figure}[htbp]
    \centering
    \includegraphics[width=0.8\textwidth]{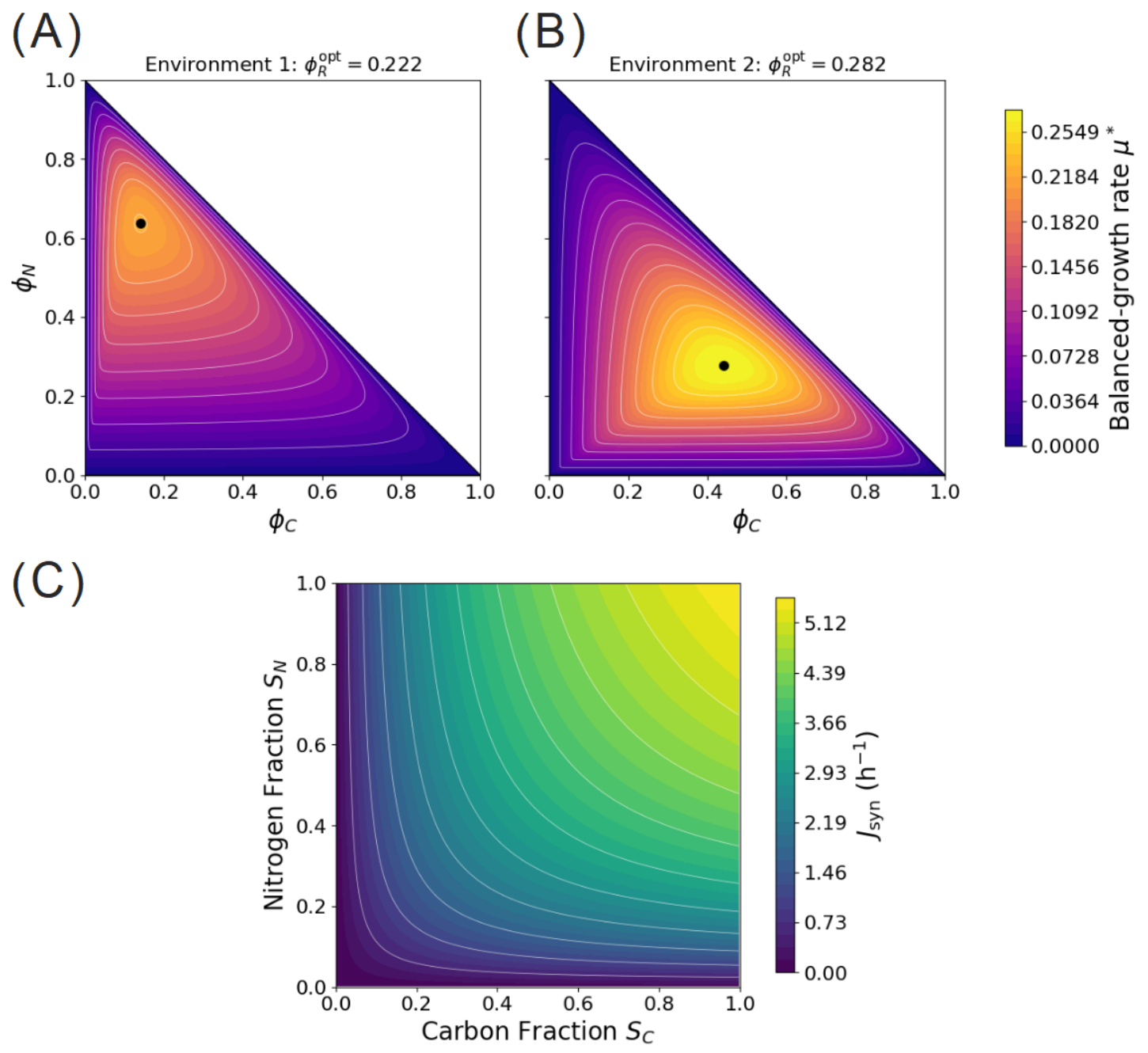}
    \caption{(A) Contour plot of steady-state growth rate $\mu^*(\phi_C,\phi_N)$ versus proteome partition $(\phi_C,\phi_N)$ for two distinct environments. A black dot indicates the optimal strategy where growth optimum is realized. The parameters used are $a=10.0\,\mathrm{h}^{-1}$, $k_C=0.4$, $k_N=0.3$, $\nu_C=0.3$, $\nu_N=0.7$. (Left) $b_C=1.0\,\mathrm{h}^{-1}$, $b_N=0.3\,\mathrm{h}^{-1}$. (Right) $b_C=0.3\,\mathrm{h}^{-1}$, $b_N=1.0\,\mathrm{h}^{-1}$. (B) Contour plot of the specific protein synthesis flux $J_{\text{syn}}$, with same parameter values as in (A).}
    \label{fig:resources}
\end{figure}

In Figure~\ref{fig:resources}(A)(B), we plot 
The resulting optimal growth rate is $\mu^{\rm opt} = 0.22\,\mathrm{h}^{-1}$ with proteome partition $(\phi_C, \phi_N, \phi_R) = (0.140, 0.638, 0.222)$ for Environment 1, and $\mu^{\rm opt} = 0.27\,\mathrm{h}^{-1}$ with proteome partition $(\phi_C, \phi_N, \phi_R) = (0.441, 0.277, 0.282)$ for Environment 2. When the environment is richer in carbon (e.g., Environment 1), more proteome is allocated to the $N$-sector; when the environment is richer in nitrogen (e.g., Environment 2), more proteome is allocated to the $C$-sector. In Figure~\ref{fig:resources}(C), the synthesis flux is shown to saturate as $S_C$ and $S_N$ increase, with the degree of saturation determined by $k_C$ and $k_N$, respectively. 

\newpage

\newpage
\section{Extended Allocation Model}
Consider the fixed system extended to $n$ functional sectors and $m$ housekeeping sectors. Denote $ r_j = R_j^{\prime} (0) $ and $ h_j = \frac{1}{2} R_j^{\prime\prime} (0) $ for $ j=1,2, \dots, n-1 $. Assuming $R_j$ can be Taylor expanded around $0$ as $ R_j (Y_j) = r_j Y_j + h_j Y_j^2 + o(Y_j^2) $, we have
\begin{subequations}
    \begin{align}
        &\frac{\mathrm{d}Y_1}{\mathrm{d}t} = bY_n - r_1 Y_1 Y_{n+1} - h_1 Y_1^2 Y_{n+1} - b Y_1 Y_n, \\
        &\frac{\mathrm{d}Y_2}{\mathrm{d}t} = r_1 Y_1 Y_{n+1} + r_1 Y_1^2 Y_{n+1}^2 - r_2 Y_2 Y_{n+2} - h_2 Y_2 Y_{n+2} - b Y_2 Y_n, \\
        &\;\;\;\vdots \notag \\
        &\frac{\mathrm{d}Y_n}{\mathrm{d}t} = \phi_1 r_{n-1} Y_{n-1} Y_{2n-1} + \phi_1 h_{n-1} Y_{n-1}^2 Y_{2n-1} - b Y_n^2, \\
        &\frac{\mathrm{d}Y_{n+1}}{\mathrm{d}t} = \phi_2 r_{n-1} Y_{n-1} Y_{2n-1} + \phi_2 h_{n-1} Y_{n-1}^2 Y_{2n-1} - b Y_n Y_{n+1}, \\
        &\;\;\;\vdots \notag \\
        &\frac{\mathrm{d}Y_{2n-1}}{\mathrm{d}t} = \phi_n r_{n-1} Y_{n-1} Y_{2n-1} + \phi_n h_{n-1} Y_{n-1}^2 Y_{2n-1} - b Y_n Y_{2n-1}, \\
        &\frac{\mathrm{d}Y_{2n}}{\mathrm{d}t} = \phi_1' r_{n-1} Y_{n-1} Y_{2n-1} + \phi_1' h_{n-1} Y_{n-1}^2 Y_{2n-1} - b Y_n Y_{2n}, \\
        &\;\;\;\vdots \notag \\
        &\frac{\mathrm{d}Y_{2n+m-1}}{\mathrm{d}t} = \phi_m' r_{n-1} Y_{n-1} Y_{2n-1} + \phi_m' h_{n-1} Y_{n-1}^2 Y_{2n-1} - b Y_n Y_{2n+m-1}.
    \end{align}
\end{subequations}

Here we use the $(n,m)=(3,0)$ case to demonstrate the proof. To emphasize the distinction from the canonical model $(2,1)$, we write $(\phi_1, \phi_2, \phi_3)$ instead of $(\phi_P, \phi_Q, \phi_R)$. It is easy to generalize to arbitrary $n$, while nonzero $m$ can be added by a factor of $q_m = 1 - \sum_{j=1}^m \phi_j'$.
\begin{subequations}
    \begin{align}
        \frac{\mathrm{d}Y_1}{\mathrm{d}t} &= b Y_3 - r_1 Y_1 Y_4 - h_1 Y_1^2 Y_4 - b Y_1 Y_3, \\
        \frac{\mathrm{d}Y_2}{\mathrm{d}t} &= r_1 Y_1 Y_4 + h_1 Y_1^2 Y_4 - r_2 Y_2 Y_5 - h_2 Y_2^2 Y_5 - b Y_2 Y_3, \\
        \frac{\mathrm{d}Y_3}{\mathrm{d}t} &= \phi_1 r_2 Y_2 Y_5 + \phi_1 h_2 Y_2^2 Y_5 - b Y_3^2, \\
        \frac{\mathrm{d}Y_4}{\mathrm{d}t} &= \phi_2 r_2 Y_2 Y_5 + \phi_2 h_2 Y_2^2 Y_5 - b Y_3 Y_4, \\
        \frac{\mathrm{d}Y_5}{\mathrm{d}t} &= \phi_3 r_2 Y_2 Y_5 + \phi_3 h_2 Y_2^2 Y_5 - b Y_3 Y_5.
    \end{align}
\end{subequations}
The unique and locally stable steady state is given by 
\begin{subequations}
    \begin{align}
        Y_1^* &= \frac{1}{2} \left(\frac{b \phi_1}{h_1 \phi_2}\right) \left[ - \left(1 + \frac{r_1 \phi_2}{b \phi_1}\right) + \sqrt{\left(1 + \frac{r_1 \phi_2}{b \phi_1}\right)^2 + \frac{4 h_1 \phi_2}{b \phi_1}} \right], \\
        Y_2^* &= \frac{1}{2} \left(\frac{b \phi_1}{h_2 \phi_3}\right) \left[ - \left(1 + \frac{r_2 \phi_3}{b \phi_1}\right) + \sqrt{\left(1 + \frac{r_2 \phi_3}{b \phi_1}\right)^2 + \frac{4 h_2 \phi_2 \phi_3}{b^2 \phi_1^2} \left(r_1 Y_1 + h_1 Y_1^2\right)} \right], \\
        Y_3^* &= \left(\frac{\phi_3}{b}\right) \left(r_2 Y_2 + h_2 Y_2^2\right), \\
        Y_4^* &= \left(\frac{\phi_2 \phi_3}{b \phi_1}\right) \left(r_2 Y_2 + h_2 Y_2^2\right), \\
        Y_5^* &= \left(\frac{\phi_3^2}{b \phi_1}\right) \left(r_2 Y_2 + h_2 Y_2^2\right).
    \end{align}
\end{subequations}

The steady-state growth rate is $\mu^* = bY_3^*$, so define the objective function $L(\phi)$ with Lagrangian multiplier $ \chi $ to be
\begin{equation}
    \begin{split}
        L(\phi) &= \left(\frac{\phi_3}{b}\right) \left(r_2 Y_2^* + h_2 (Y_2^*)^2\right) - \chi \left(\phi_1 + \phi_2 + \phi_3 - 1\right) \\
        &= b \phi_1 - b^2 \phi_1^{2}\!\left(\frac{1}{r_1\phi_2}+\frac{1}{r_2\phi_3}\right) \\
        &+ b^{3} \phi_1^3 \left[
        \frac{1}{r_1^{2}\phi_2^{2}}
        +\frac{h_1}{r_1^{3}\phi_2^{2}}
        +\frac{1}{r_1r_2\phi_2\phi_3}
        +\frac{1}{r_2^{2}\phi_3^{2}}
        +\frac{h_2}{r_2^{3}\phi_3^{2}}
        \right] \\
        &- \chi \left(\phi_1 + \phi_2 + \phi_3 - 1\right) +O((b/r_2)^4).    
    \end{split}
\end{equation}
Since there are many competing timescales $r_j(Y_j)$, we rescale by the rate of the synthesis flux $r_{n-1}$. 
The partial derivatives are
\begin{subequations}
    \begin{align}
        &\frac{\partial L}{\partial \phi_1} = b - 2 b^2 \phi_1 \!\left(\frac{1}{r_1\phi_2}+\frac{1}{r_2\phi_3}\right) - \eta + O\!\left((b/r_2)^3\right)=0, \\
        &\frac{\partial L}{\partial \phi_j} = \frac{b^2\phi_1^2}{r_1 \phi_j^2} - \eta + O\!\left((b/r_2)^3\right)=0, \quad j=2,3.
    \end{align}
\end{subequations}
From the last two equations, $\phi_2 \colon \phi_3 = \frac{1}{\sqrt{r_1}} \colon \frac{1}{\sqrt{r_2}}$. Eliminating $\chi$ gives a quadratic equation in the ratio $\phi_1/\phi_3$, up to order $(b/r_2)^3$: 
\begin{equation}
    \left(\frac{\sqrt{r_2} \phi_3}{\phi_1}\right)^2 - 2b \left(\frac{1}{\sqrt{r_1}} + \frac{1}{\sqrt{r_2}}\right) \left(\frac{\sqrt{r_2} \phi_3}{\phi_1}\right) - b = 0.
\end{equation}
Then, letting $\mathcal{A} = \frac{1}{\sqrt{r_1}} + \frac{1}{\sqrt{r_2}}$, we have
\begin{equation}
    \frac{\sqrt{r_2} \phi_3}{\phi_1} = b\mathcal{A} + \sqrt{b^2 \mathcal{A}^2 + b}
\end{equation}
Finally, the normalization condition $\phi_1 + \phi_2 + \phi_3 = 1$ gives, to leading orders,
\begin{subequations}
    \begin{align}
        \phi_1^{\text{opt}} &\simeq 1 - \mathcal{A}\sqrt{b} + O\!\left(b/r_2\right), \\
        \phi_j^{\text{opt}} &\simeq \frac{1}{\sqrt{r_{j-1}}} \sqrt{b} + O\!\left(b/r_2\right), \quad j=2,3.
    \end{align}
\end{subequations}

In the general model with $(n,m)$ sector, the objective up to $O (\tilde{b}^3) $ is of the form 
\begin{equation}
    L(\phi, \chi) = \left(1 - \sum_{k=1}^m \phi'_k\right)
    \left[b \phi_1 - b^2 \phi_1^2 \left(\sum_{j=1}^{n-1} \frac{1}{r_j \phi_{j+1}}\right)\right] 
    - \chi \left(\sum_{j=1}^n \phi_j + \sum_{k=1}^m \phi'_k- 1\right). 
\end{equation}

Similarly, the non-transporter proteome partition fractions satisfy
\begin{equation}
    \phi_2 \colon \phi_3 \colon \cdots \colon \phi_n = \frac{1}{\sqrt{r_1}} \colon \frac{1}{\sqrt{r_2}} \colon \cdots \colon \frac{1}{\sqrt{r_{n-1}}}. 
\end{equation}
Following the same procedure, assuming all $r_j$ are of the same order, we get  
\begin{subequations}
    \begin{align}
        \phi_1^{\text{opt}} &\simeq \left(1 - \sum_{k=1}^m \phi'_k\right)\left(1 - \mathcal{A}\sqrt{b}\right) + O\!\left(b/r_{n-1}\right), \\
        \phi_j^{\text{opt}} &\simeq \left(1 - \sum_{k=1}^m \phi'_k\right)\left(\frac{1}{\sqrt{r_{j-1}}}\sqrt{b}\right) + O\!\left(b/r_{n-1}\right), 
        \quad j=2,...,n.
    \end{align}
\end{subequations}

\end{document}